\documentclass[journal,draftclsnofoot,onecolumn,12pt]{IEEEtran}

\usepackage{amsmath}
\usepackage{amssymb}
\usepackage[dvips]{graphicx}
\graphicspath{{./Figures/}}
\usepackage{curves,color}
\usepackage{subfigure}
\usepackage{dsfont,slashbox,wasysym}
\usepackage{tikz}
\usepackage{pdfpages,enumerate}
\usepackage{multirow}
\usepackage{arydshln}
\usepackage{algorithmic}
\usepackage{algorithm}

\allowdisplaybreaks

\DeclareMathOperator{\E}{\mathbb{E}}
\newcommand{\argmin}{\mathop{\mathrm{argmin}}}      
\newcommand{\vect}[1]{\boldsymbol{#1}}

\RequirePackage[normalem]{ulem} 
\RequirePackage{color}\definecolor{RED}{rgb}{1,0,0}\definecolor{BLUE}{rgb}{0,0,1} 

\begin{document}
\title{Design of Irregular SC-LDPC Codes With Non-Uniform Degree Distributions by Linear Programing}

\author{Heeyoul~Kwak, Jong-Seon~No, \IEEEmembership{Fellow,~IEEE}, and~Hosung~Park, \IEEEmembership{Member,~IEEE} 
\thanks{H.~Kwak and J.-S.~No are with the Department of Electrical and Computer Engineering, INMC, Seoul National University, Seoul 08826, Korea (e-mail: ghy1228@ccl.snu.ac.kr, jsno@snu.ac.kr).}
\thanks{H.~Park is with the School of Electronics and Computer Engineering, Chonnam National University, Gwangju 61186, Korea (e-mail: hpark1@jnu.ac.kr).}}

\maketitle

\begin{abstract}
In this paper, we propose a new design method of irregular spatially-coupled low-density parity-check (SC-LDPC) codes with non-uniform degree distributions by linear programming (LP). In general, irregular SC-LDPC codes with non-uniform degree distributions is difficult to design with low complexity because their density evolution equations are multi-dimensional. To solve the problem, the proposed method is based on two main ideas: A local design of the degree distributions and pre-computation of the input/output message relationship. These ideas make it possible to design the degree distributions of irregular SC-LDPC codes by solving low complexity LP problems over the binary erasure channel. We also find a proper objective function for the proposed design methodology to improve the performance of SC-LDPC codes. It is shown that the irregular SC-LDPC codes obtained by the proposed method are superior to regular SC-LDPC codes in terms of both asymptotic and finite-length performances. 
\end{abstract}

\begin{IEEEkeywords}
Linear programming (LP) problem, low-density parity-check (LDPC) codes, non-uniform degree distributions, spatially-coupled low-density parity-check (SC-LDPC) codes.
\end{IEEEkeywords}

\section{Introduction}
\IEEEPARstart{S}{patially}-coupled low-density parity-check (SC-LDPC) codes have attracted much attention due to their ability to universally achieve channel capacity over general binary memoryless symmetric (BMS) channels under iterative belief propagation (BP) decoding \cite{Kudekar}. By coupling $L$ disjoint component low-density parity-check (LDPC) codes arranged at $L$ positions, SC-LDPC codes are constructed and their decoding performances under BP decoding approaches the maximum a posteriori (MAP) decoding performance of uncoupled LDPC codes. This phenomenon is termed the {\em threshold saturation effect}, which is empirically observed in \cite{Lentmaier} and analytically proved for the binary erasure channel (BEC) \cite{Kudekar1} and general BMS channels \cite{Kudekar}, \cite{Kudekar2}. 

Most studies of SC-LDPC codes focus on spatially-coupled codes composed of regular LDPC codes. However, irregular LDPC codes can be coupled to construct irregular SC-LDPC codes, where the threshold saturation effect also occurs \cite{Kudekar}, \cite{Yedla}, \cite{Yedla2}. Thus, there have been researches to construct SC-LDPC codes from irregular LDPC codes which outperform regular LDPC codes. The first approach to construct irregular SC-LDPC codes is coupling protograph-based codes \cite{Thorpe} such as repeat-accumulate (RA) \cite{Divsalar}, accumulate-repeat-jagged-accumulate (ARJA) \cite{Divsalar2}, and MacKay-Neal (MN) \cite{Mackay} codes. Spatially coupled ARJA codes \cite{Mitchell1}, \cite{Mitchell} and spatially-coupled MN codes \cite{Kasai}, \cite{Obata} show better asymptotic performance than regular SC-LDPC codes with bounded degrees. However, it is later known that the finite-length performance of spatially-coupled ARJA codes is rather worse than that of other spatially-coupled codes due to their inferior scaling behavior \cite{Stinner}. On the contrary, spatially-coupled repeat accumulate (SC-RA) codes \cite{Johnson} show the best finite-length performance compared to other spatially-coupled codes \cite{Stinner}. 

Similar to protograph-based spatially-coupled codes, randomly constructed irregular SC-LDPC codes can be constructed by coupling irregular LDPC codes with variable and check node degree distributions $\lambda(x)$ and $\rho(x)$. However, it is difficult to globally optimize the degree distributions $\lambda(x)$ and $\rho(x)$ of irregular SC-LDPC codes with low-complexity because their density evolution (DE) equations \cite{Kudekar1} are multi-dimensional for the number of positions $L$. Thus, many studies assume a nearly regular degree distribution with two distinct degrees because its optimization is possible by exhaustive searches. For example, SC-LDPC ensembles with two distinct degrees on check nodes are designed to improve the BP threshold over wide range of code rate \cite{Nitzold}. In \cite{Aref}, they show that slightly imposing irregularity on the variable node degree distribution $\lambda(x)$ can improve the convergence speed. Also, it has been reported that this improvement in the convergence speed also results in an improvement of the finite-length performance \cite{Schmalen}. Recently, an optimization method for irregular SC-LDPC codes without any constraints on the degree distributions is proposed \cite{Koganei}. Further, SC-LDPC ensembles with so called {\em non-uniform degree distributions} are designed, where the degree distributions can differ for each position. By optimizing the degree distributions using a genetic algorithm, it is shown that the optimized SC-LDPC codes exhibit improved decoding performance compared to regular SC-LDPC codes.

In this paper, motivated by the ideas in \cite{Koganei}, we propose a systematic method with low-complexity to design irregular SC-LDPC codes with non-uniform degree distributions. To reduce the design complexity, the proposed method iteratively solves linear programming (LP) problems. The LP problem is widely used when optimizing uncoupled irregular LDPC codes over the BEC \cite{Richardson}, \cite{Jamali}. However, it is hard to directly apply the LP problem to the designing of the degree distributions of SC-LDPC codes because the DE equations are multi-dimensional. Thus, we introduce two important methods for designing the degree distributions. First, instead of designing the degree distributions at once, local designs are iteratively performed, where the variable node degree distributions of the target positions are obtained one at a time while keeping the degree distributions of the other positions constant. Second, the input/output message relationship between the variable nodes at the target positions and the remaining graph is pre-computed. These two methods allow us to obtain a {\em one-dimensional DE equation}, which enables the designing of the degree distributions simply, such as by solving LP problems. 

The one-dimensional DE equation is used in \cite{Sanatkar} to optimize the degree distribution of additional variable nodes to mitigate the rate-loss of SC-LDPC codes, where the objective function for the optimization is to maximize the design rate. However, we observe that the objective function of maximizing the design rate is not proper when designing the degree distributions of irregular SC-LDPC codes.
Thus, we use another objective function which minimizes the number of required iterations for successful decoding \cite{Jamali}, \cite{Smith}. The proposed design method finds the degree distributions of the target positions one at a time to minimize the number of required iterations in the one-dimensional DE equation. Although the proposed method aims to minimize the number of required iterations in the one-dimensional DE equation, it is observed that the overall number of required iterations in the multi-dimensional DE equations is also properly decreased by the proposed design method. 
In addition, the BP threshold is improved by the proposed design method if all of the degree distributions are locally designed, which is confirmed graphically from an analysis of the expected graph evolution \cite{Olmos}. Numerical results show that irregular SC-LDPC codes with the degree distributions obtained by the proposed method exhibit superior BP threhold and finite-length performance compared to regular SC-LDPC codes. It is also shown that an additional degree of freedom on the check node degrees can improve the performance further. Finally, we introduce several methods which can be used to improve the performance by adopting the multi-edge type (MET) structure \cite{Richardson} and applying the proposed method to SC-RA codes.

The remainder of the paper is organized as follows. Section II introduces the construction methods
of SC-LDPC ensembles and the tools for analyzing the ensembles. In Section III, the methods to design the degree distributions of SC-LDPC ensembles are proposed. In Section IV, the finite-length performance of the SC-LDPC codes obtained by the proposed methods is shown and several methods for improving the performance further are presented. Finally, the conclusion is given in Section V.

\section{SC-LDPC Ensembles}
\subsection{Construction of SC-LDPC Ensembles}\label{Sec:code_construction}
Let $l$ and $r$ denote the variable and check node degrees of regular LDPC ensembles, respectively. The SC-LDPC ensemble considered in this paper is irregular SC-LDPC ensembles, where the degree distributions of the variable nodes at each position can be irregular while the check node degrees at each position are regular but can differ at each position. The SC-LDPC ensemble consists of $M$ variable nodes located at each of $L$ positions with $M_c$ check nodes located at each of $L+w-1$ positions, where $M_c\triangleq M\frac{l}{r}\in\mathbb{N}$. The variable nodes at position $u$ follow the edge perspective degree distribution $\lambda_{u}(x)=\sum_d\lambda_{u,d}x^{d-1}$ \cite{Richardson} while the check nodes at position $v$ have degree $r_v\in\mathbb{N}$. The number of edges between the positions in the graph is represented by a $(L+w-1)\times L$ connectivity matrix $\vect{T}$, where the entry $T_{v,u}$ is the number of edges between the check nodes at position $v$ and the variable nodes at position $u$. For example, the regular SC-LDPC ensemble \cite{Kudekar1} constructed by coupling regular LDPC ensembles has code parameters of $\lambda_{u}(x)=x^{l-1}$ for $1\le u\le L$, $r_{v}=r$ for $1\le v \le L+w-1$, and ${T}_{v,u}={Ml}/{w}$ for $u\le v \le u+w-1$. 

\begin{figure*}[t]
\centering
\subfigure[Graph 1]{\includegraphics[scale=0.45]{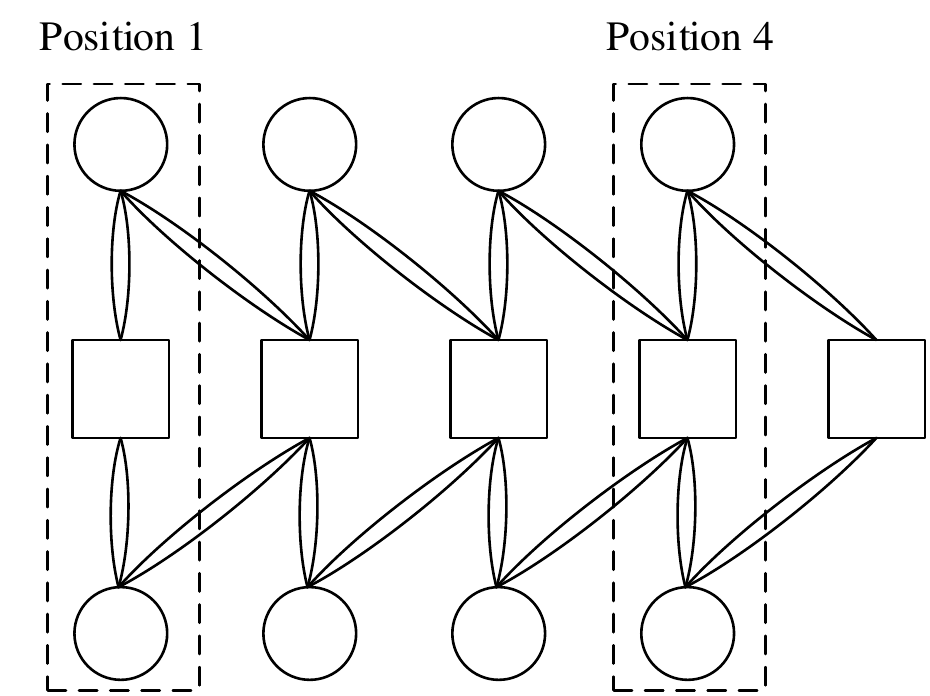}}%
\hspace{10pt}
\subfigure[Graph 2]{\includegraphics[scale=0.45]{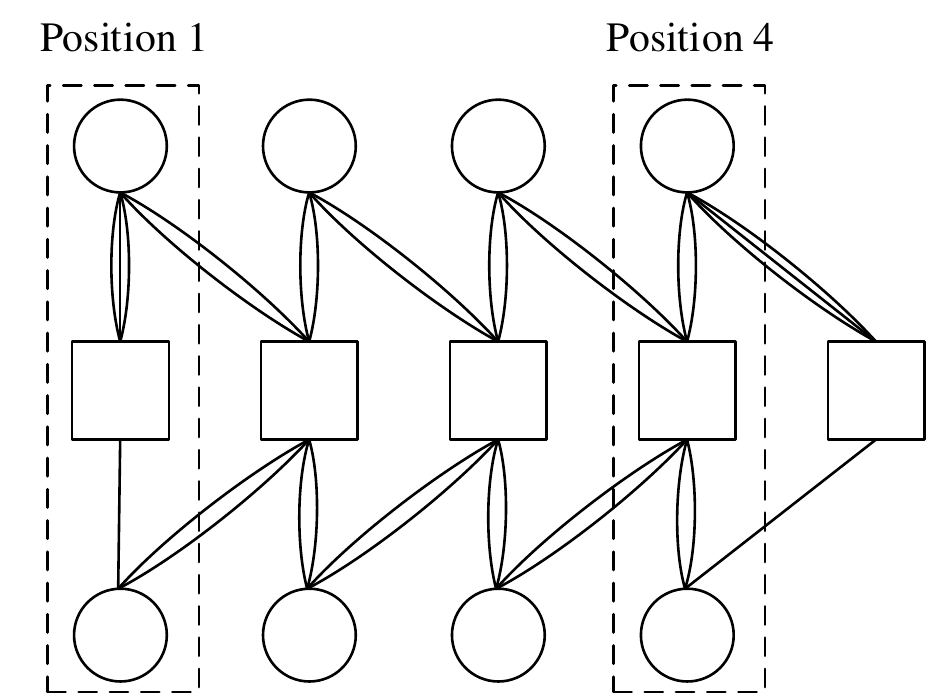}}%
\hspace{10pt}
\subfigure[Graph 3]{\includegraphics[scale=0.45]{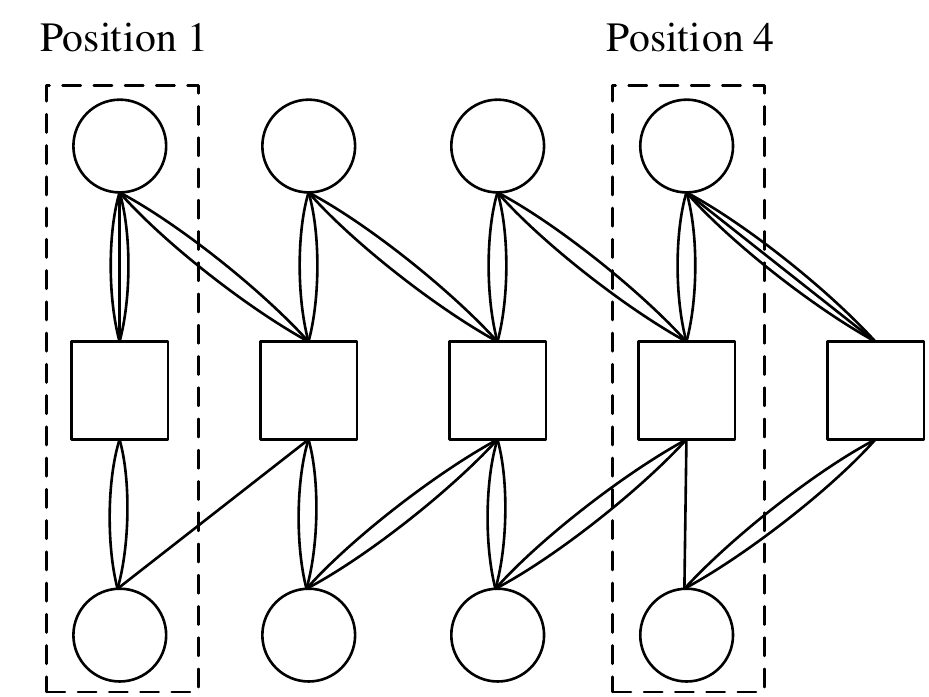}}%
\caption{Examples of Tanner graphs of SC-LDPC codes for $L=4$, $w=2$, and $M=2$.}
\label{Fig:Example}
\end{figure*}

In Fig.~\ref{Fig:Example}, Tanner graphs of three different SC-LDPC codes with $L=4, w=2$, and $M=2$ are shown as examples and their code parameters are represented in Table \ref{Table:Example}. Note that there are multiple edges in the Tanner graphs to represent various degree distributions with a small number of nodes. First, Graph 1 represents a regular SC-LDPC code, where the degrees of all of the variable nodes are equal to $4$. In contrast, the degree distributions of the variable nodes at positions $1$ and $4$ in Graph $2$ are irregular. Graph 2 is called a SC-LDPC code with non-uniform degree distributions in that the degree distributions of the variable nodes at each position differ from each other. The objective of this paper is to find good degree distributions of the variable nodes at each position of the SC-LDPC ensemble. In addition, Graph 2 becomes Graph 3 after permitting a different number of connected edges between the positions, giving more freedom in the design of the degree distributions. Note that we assume a symmetric structure, that is, $\lambda_{u}(x)=\lambda_{L+1-u}(x), r_{v}=r_{L+w-v}, {T}_{v,u}=T_{L+w-v,L+1-u}$ for all the codes in this paper.

\begin{table}[t]
\centering
\caption{Parameters of the Tanner graphs in Fig. 1}
\label{Table:Example}
\begin{tabular}{c|l|c}\hline
                        & \multicolumn{1}{c|}{$\lambda_{i}(x)$} & ${\vect T}$         \\ \hline
\multirow{3}{*}{Graph 1} & $\lambda_{1}(x)=x^3$                  & \multirow{3}{*}{$\arraycolsep=2pt\def\arraystretch{0.8}\left[\begin{array}{cccc}
4 &  & &\\
4 & 4 & &\\
  & 4 &4 &\\
  &  & 4 & 4\\
 & & & 4
\end{array}\right]$} \\
                        & $\lambda_{2}(x)=x^3$                  &                   \\
                        & $\lambda_{3}(x)=x^3$                  &                   \\ 
                        & $\lambda_{4}(x)=x^3$                  &                   \\ \hline
\multirow{3}{*}{Graph 2} & $\lambda_{1}(x)=\frac{3}{8}x^2+\frac{5}{8}x^4$                & \multirow{3}{*}{$\arraycolsep=2pt\def\arraystretch{0.8}\left[\begin{array}{cccc}
4 &  & &\\
4 & 4 & &\\
  & 4 &4 &\\
  &  & 4 & 4\\
 & & & 4
\end{array}\right]$} \\
                        & $\lambda_{2}(x)=x^3$                  &                   \\
                        & $\lambda_{3}(x)=x^3$                &                   \\ 
                        & $\lambda_{4}(x)=\frac{3}{8}x^2+\frac{5}{8}x^4$                 &                   \\ \hline
\multirow{3}{*}{Graph 3} & $\lambda_{1}(x)=\frac{3}{8}x^2+\frac{5}{8}x^4$                & \multirow{3}{*}{$\arraycolsep=2pt\def\arraystretch{0.8}\left[\begin{array}{cccc}
5 &  & &\\
3 & 4 & &\\
  & 4 &4 &\\
  &  & 4 & 3\\
 & & & 5
\end{array}\right]$} \\
                        & $\lambda_{2}(x)=x^3$                  &                   \\
                        & $\lambda_{3}(x)=x^3$                &               \\
                        & $\lambda_{4}(x)=\frac{3}{8}x^2+\frac{5}{8}x^4$                &               \\\hline
\end{tabular}   
\end{table}

For the given code parameters $\lambda_{u}(x)$, $r_v$, and ${\vect T}$, the detailed construction method of the SC-LDPC ensemble is described as follows. Let $[m,n]\triangleq\{m,m+1,\ldots,n\}$ for $m<n$. First, place $M$ variable nodes with degree distribution $\lambda_{u}(x)$ at position $u$ for $u\in[1,L]$. Then, there are $M/{\int_{0}^{1}\lambda_{u}(x) dx}$ variable node sockets at position $u$, where ${1}/{\int_{0}^{1}\lambda_{u}(x) dx}$ is the average variable node degree of the variable nodes at position $u$. To connect all of the sockets, the number of edges connected to the variable nodes at position $u$ should be equal to the number of sockets, that is, $M/{\int_{0}^{1}\lambda_{u}(x) dx}=\sum_{i}T_{i,u}$. Let $\pi_{u}$ be a random permutation on $[1,\sum_{i}T_{i,u}]$ for position $u$. Divide $\pi_{u}$ into $w$ disjoint subsets denoted by $\pi_{u}^{1},\ldots,\pi_{u}^{w}$ such that the size of $\pi_{u}^{t}$ becomes $T_{u+t-1,u}$ for $t\in[1,w]$. Similarly, place $M_c$ check nodes of degree $r_v$ at position $v$ for $v\in[1,L+w-1]$ in the graph. Accordingly, there are $M_c r_{v}$ check node sockets at position $v$ but some sockets cannot be filled if $\sum_{j}T_{v,j}<M_c r_v$. Divide the randomly selected $\sum_{j}T_{v,j}$ check node sockets among the $M_c r_v$ check node sockets at position $v$ into $w$ groups randomly such that the size of group $t$ becomes $T_{v,v-t+1}$ for $t\in[1,w]$, where $T_{v,u}=0$ for $u<0$. Then, the $j$th check node socket in group $t$ at position $u+t-1$ is connected to the $\pi_{u}^{t}(j)$th variable node socket at position $u$, where $\pi_{u}^{t}(j)$ denotes the $j$th element of $\pi_u^{t}$.

The total numbers of variable nodes $V$ and check nodes $C$ placed in the graph are $V=LM$ and $C=(L+w-1)M_c$, respectively. 
However, some check nodes at position $v$ for $v\in[1,w]$ or $v\in[L+1,L+w-1]$ cannot be connected to any variable nodes in the graph because the number $\sum_{j}T_{v,j}$ of edges connected to check nodes is lower than the number of sockets $M_c r_v$. To be specific, a check node socket at position $v$ for $v\in[1,w]$ or $v\in[L+1,L+w-1]$ cannot be connected to any variable node in the graph with probability $1-{\sum_{j}T_{v,j}}/(M_c r_v)$. Then, the expected number of check nodes $\overline{C}$ with at least one connection to the variable nodes in the graph is given as 
\begin{equation*}
\overline{C}=\left(L+w-1-2\sum_{v=1}^{w-1}\left(1-\frac{\sum_{j}T_{v,j}}{M_c r_v}\right)^{r_v}\right)M_c,
\end{equation*}
 which is less than $C=(L+w-1)M_c$. Therefore, the design rate $R_{\rm SC}=1-\overline{C}/V$ of the SC-LDPC ensemble is expressed as
\begin{align}
R_{\rm SC}=\bigg(1-\frac{l}{r}\bigg)-\frac{l}{r}\frac{w-1}{L}+\frac{l}{r}\frac{2\sum\limits_{v=1}^{w-1}\left(1-\frac{\sum_{j}T_{v,j}}{M_c r_v}\right)^{r_v}}{L}. \label{Eq:Design_rate}
\end{align}
Generally, because the last term of (\ref{Eq:Design_rate}) is much smaller than the other terms, we ignore the last term when calculating the design rate for convenience.

\subsection{DE Equations of SC-LDPC Ensembles}
In this paper, the channel is assumed to be the BEC with erasure probability $\epsilon$. 
The DE is often used in predicting the performance in an asymptotic setting such as infinite code lengths and infinite numbers of iterations. Let $x_u^{(\ell)}$ and $y_v^{(\ell)}$ denote the average erasure probabilities of messages at iteration $\ell$ emitted from the variable nodes at position $u$ and the check nodes at position $v$, respectively. Set the initial conditions as $x_u^{(0)}=\epsilon$ for all $u$. Then, the evolution of $x_u^{(\ell)}$ can be expressed as

\begin{align}
y_{v}^{(\ell)}&=1-\left(1-\frac{\sum_{j}T_{v,j}x_{j}^{(\ell)}}{M_c r_v}\right)^{r_v-1},
x_{u}^{(\ell+1)}=\epsilon\lambda_{u}\left(\frac{\sum_{i}T_{i,u}y_{i}^{(\ell)}}{\sum_{i}T_{i,u}}\right).\label{Eq:DE}
\end{align}
With the DE equations in (\ref{Eq:DE}), we can obtain the BP threshold which is defined as the maximum $\epsilon$ for which $x_u^{(\ell)}$ goes to zero for all $u$. Additionally, we define the overall number of required iterations $I_{r}$ for successful decoding as the minimum $\ell$ such that $x_u^{(\ell)}=0$ for all $u$ and define the average convergence speed as $L/I_{r}$ \cite{Aref}.

\subsection{Expected Graph Evolution of SC-LDPC Ensembles}\label{Sec:EGE}
In \cite{Stinner}, \cite{Olmos}, the scaling law of SC-LDPC codes is derived to predict the finite-length performance of SC-LDPC codes. The scaling law of SC-LDPC ensembles depends on the scaling parameters derived by analyzing the statistical behavior of the number of degree-one check nodes in the remaining graph of the peeling decoding. 
In \cite{Olmos}, a system of coupled differential equations to compute the expected number of degree-one check nodes is derived for regular SC-LDPC ensembles. Because the SC-LDPC ensemble considered in this paper has the non-uniform degree distributions, the differential equations to compute the expected number of degree-one check nodes should be modified as follows.

Consider the degree distributions of the original graph before the peeling decoder is initialized. First, consider the degree distribution of the variable nodes at position $u$. Define the type of a variable node at position $u$ using the vector ${\underline x}=({ x}_1,\ldots,{ x}_{w})$, where ${ x}_t$ represents the number of edges connected to the check nodes at position $u+t-1$. Although the profile of degrees of each variable node is sufficient to represent the DE equations in (\ref{Eq:DE}), the type of each variable node should be considered when calculating the expected graph evolution. Let $\mathbb{P}_{u}(\underline x)$ be the probability that a variable node chosen at random from position $u$ in the original graph is of type ${\underline x}$. Considering the construction method of the SC-LDPC ensemble, we can see that the vector ${\underline x}=({ x}_1,\ldots,{ x}_{w})$ for the variable nodes at position $u$ follows a multinomial distribution with probabilities ${\underline p}=({p}_1,\ldots,{p}_{w})$, where ${ p}_t={T_{u+t-1,u}}/{\sum_i T_{i,u}}$.
Thus, the probability $\mathbb{P}_{u}(\underline x)$ is given as 
\begin{equation*}
\mathbb{P}_{u}({\underline x})=\frac{|{\underline x}|!}{{x}_1!\cdots {x}_{w}!}\prod_{t=1}^{w}\left(\frac{T_{u+t-1,u}}{\sum_i T_{i,u}}\right)^{x_t}
\end{equation*}
where $|{\underline x}|=\sum_{i}{x}_i$. Next, consider the check nodes at position $v$.  Let $\rho_{m,v}$ be the probability that a check node chosen at random from position $v$ in the original graph is of degree $m$, which is given as 
\begin{equation*}
\rho_{m,v}=\begin{cases}\binom{r_v}{m}\Big(\frac{\sum_{j}T_{v,j}}{M_cr_v}\Big)^{m}\Big(1-\frac{\sum_{j}T_{v,j}}{M_c r_v}\Big)^{r_v-m}, {\rm~if~} v\in[1,w-1]\\
1,{\rm~if~} m=r_v, v\in[w,L]\\
0,{\rm~if~} m<r_v, v\in[w,L]\\
\rho_{m,L+w-v},{\rm~if~} v\in[L+1,L+w-1].
\end{cases}
\end{equation*}

For iteration $\ell$ of the peeling decoder, let $\tau$ be the number of iterations normalized by $M$, i.e., $\tau=\ell/M$. At time $\tau$, let $R_{j,v}(\tau)$ be the number of edges connected to the check nodes of degree $j$, $j=1,\ldots,r_v$, at position $v$, $v\in[1,L+w-1]$. Likewise, let $U_{{\underline x},u}(\tau)$ be the number of edges that are connected to variable nodes of type ${\underline x}$ at position $u$, $u\in[1,L]$.
 After the initialization of the peeling decoder, the expected value of $R_{j,v}(0)$ is expressed as

\begin{equation*}
\E[R_{j,v}(0)]=jM_c\sum_{m\ge j}^{r_v}\rho_{m,v}\binom{m}{j}\epsilon^{j}(1-\epsilon)^{m-j}.
\end{equation*}

In addition, the initial values of the expected value of $U_{{\underline x},u}(\ell)$ can be computed as
\begin{equation*}
\E[U_{{\underline x},u}(0)]=\begin{cases}\epsilon \lambda_{u,|{\underline x}|}\frac{M}{\int_{0}^{1}\lambda_{u}(z)dz}\mathbb{P}_{u}({\underline x}),&u\in[1,L]\\
0,&{\rm otherwise}.
\end{cases}
\end{equation*}

Let $\E[\Delta R_{j,v}(\tau)]=\E[R_{j,v}(\tau+1/M)-R_{j,v}(\tau)]$ and $\E[\Delta U_{{\underline x},u}(\tau)]=\E[U_{{\underline x},u}(\tau+1/M)-U_{{\underline x},u}(\tau)]$, where the expectation is determined given the degree distributions in the remaining graph at time $\tau$. In order to compute the expectation of $R_{j,v}(\tau)$ and $U_{{\underline x},u}(\tau)$ from the initial values, we solve the system of differential equations described as
\begin{align*}
\frac{\partial R_{j,v}(\tau)}{\partial \tau}=\frac{\E\left[\Delta R_{j,v}(\tau)\right]}{1/M},
 \frac{\partial U_{{\underline x},u}(\tau)}{\partial \tau}=\frac{\E\left[\Delta U_{{\underline x},u}(\tau)\right]}{1/M}.
\end{align*}
 
The procedure used to obtain $\E[\Delta R_{j,v}(\tau)]$ and $\E[\Delta U_{{\underline x},u}(\tau)]$ is described as follows. Let $\phi_{m,{\underline x},u}(\tau)$ be the probability that a variable node of type ${\underline x}$ connected to a degree-one check node at position $m$ belongs to position $u$. Then, we have 
\begin{equation*}
\phi_{m,{\underline x},u}(\tau)=\begin{cases}\frac{\frac{{x}_{m-u+1}}{|{\underline x}|}U_{{\underline x},u}(\tau)}{\sum\limits_{i\in S(m)}\left(\sum_{{\underline x}'}\frac{{ x}'_{m-i+1}}{|{\underline x}'|}U_{{\underline x}',i}(\tau)\right)},& {\rm if~}u\in S(m)\\
0,&{\rm otherwise~}
\end{cases}
\end{equation*}
where $S(m)=\{j|\min(m-(w-1),1)\le j \le m\}$. When a degree-one check node from position $m$ and the variable node connected to it are removed, we define $\xi_{m,v,t}(\tau)$ as the probability that $t$ edges of the removed variable node are connected to the check nodes other than the removed check node at position $v$. Then, we have
\begin{align*}
\xi_{m,v,t}(\tau)=\begin{cases}
\sum\limits_{i\in S(v)}\bigg(\sum\limits_{{\underline x}: {x}_{v-i+1}=t}\phi_{m,{\underline x},i}(\tau)\bigg),&{\rm if~} m\neq v\\
\sum\limits_{i\in S(v)}\bigg(\sum\limits_{{\underline x}: { x}_{v-i+1}=t+1}\phi_{m,{\underline x},i}(\tau)\bigg),&{\rm if~} m=v\
\end{cases}
\end{align*}
for $t\le l_{\max}-1$, where $l_{\max}$ denotes the maximum degree of variable nodes. The average number of degree-$j$ check nodes losing one edge when $t$ edges are randomly removed from check nodes at position $v$ is given as
\begin{equation*}
F_{j,v,t}(\tau)=\sum_{k=1}^{t}k\binom{t}{k}\delta_{j,v}^{k}(\tau)\left(1-\delta_{j,v}(\tau)\right)^{t-k}
\end{equation*}
where $\delta_{j,v}(\tau)={R_{j,v}(\tau)}/{\sum\limits_{q=1}^{r_v}R_{q,v}(\tau)}$ for $j\le r_v$ and $\delta_{r_v+1,v}(\tau)=0$. Then, we have
\begin{align*}
&\E[\Delta U_{{\underline x},u}(\tau)|{\rm pos}(\tau)=m]=-|{\underline x}|\phi_{m,{\underline x},u}(\tau)\label{Eq:Evolution_U}\\
&\E[\Delta R_{j,v}(\tau)|{\rm pos}(\tau)=m]=\nonumber
\begin{cases}
j\sum\limits_{t=1}^{l_{\max}-1}\xi_{m,v,t}(\tau)\bigg(F_{j+1,v,t}(\tau)-F_{j,v,t}(\tau)\bigg)-1,&{\rm if~}v=m, j=1\\
j\sum\limits_{t=1}^{l_{\max}-1}\xi_{m,v,t}(\tau)\bigg(F_{j+1,v,t}(\tau)-F_{j,v,t}(\tau)\bigg),&{\rm otherwise}
\end{cases}
\end{align*}
where ${\rm pos}(\tau)$ is the position at which a degree-one check node is removed at time $\tau$. Finally, the expectations of $\Delta R_{j,v}(\tau)$ and $\Delta U_{{\underline x},u}(\tau)$ are described as
\begin{align*}
&\E[\Delta R_{j,v}(\tau)]=\sum_{m=1}^{L+w-1}\E[\Delta R_{j,v}(\tau)|{\rm pos}(\tau)=m]p_m(\tau)\\
&\E[\Delta U_{{\underline x},u}(\tau)]=\sum_{m=1}^{L+w-1}\E[\Delta U_{{\underline x},u}(\tau)|{\rm pos}(\tau)=m]p_m(\tau)
\end{align*}
where $p_m(\tau)={R_{1,m}(\tau)}/{\sum\limits_{v=1}^{L+w-1}R_{1,v}(\tau)}.$ Then, the expectation of the total number of degree-one check nodes $\E[R_{1}(\tau)]$ in the remaining graph at $\tau$ becomes $\sum_{v}\E[R_{1,v}(\tau)]$ and the normalized number of degree-one check nodes $r_{1}(\tau)$ is computed as $r_1(\tau)=\E[R_{1}(\tau)]/M$.

\section{Proposed Design Methods of SC-LDPC Ensembles}
In this section, we propose new design methods of code parameters $\lambda_{u}(x)$, $\rho_v$, and $T_{v,u}$ for the SC-LDPC ensembles. 

\subsection{Pre-Computation of $\delta_{u}(z)$}\label{Sec:pre-computation}
Before describing the proposed design methods of the SC-LDPC ensemble, we introduce a procedure called pre-computation of $\delta_{u}(z)$ and described in  Algorithm \ref{alg:Pre-computation of delta}. This procedure, firstly used in \cite{Sanatkar}, is essential for the proposed design methods. 
\begin{algorithm}[h]
 \caption{Pre-computation of $\delta_{u}(z)$ \cite{Sanatkar}}
 \label{alg:Pre-computation of delta}
 \begin{algorithmic}[1]
 \renewcommand{\algorithmicrequire}{\textbf{Input:} }
 \REQUIRE $\{\lambda_{1}(x),\ldots,\lambda_{L}(x)\}, \{r_1,\ldots,r_{L+w-1}\},{\vect T},Q,\epsilon,u$
 \FOR {$q=1:Q$}
 \STATE Set $z_q=\epsilon\frac{q}{Q}$
\STATE Update $x_i^{(\ell)}$ for all $i$ except $u$ and $L-u+1$ by DE equations (\ref{Eq:DE}) until the messages are saturated while fixing $x_{u}^{(\ell)}=z_q, x_{L-u+1}^{(\ell)}=z_q$ for all $\ell$.
\STATE Let $\delta_{u}(z_q)$ be the incoming message to variable nodes at position $u$, that is,
\begin{equation*}
\delta_{u}(z_q)=\frac{\sum_{i}T_{i,u}y_{i}^{(\ell)}}{\sum_{i}T_{i,u}}.
\end{equation*}
\ENDFOR
 \end{algorithmic}
 \end{algorithm}
 
In Algorithm \ref{alg:Pre-computation of delta}, the graph can be considered as two parts, which are the variable nodes at positions $u$ and $L-u+1$ and the remaining graph, and then the input/output message relationship is obtained. First, the input message $z$ to the remaining graph is passed from the variable nodes at positions $u$ and $L-u+1$ to the remaining graph. And then the messages in the remaining graph are updated using the DE equations (\ref{Eq:DE}) until all of the messages are saturated while the messages from the variable nodes at positions $u$ and $L-u+1$ to the remaining graph are fixed at $z$. Finally, let the incoming message to the variable nodes at positions $u$ and $L-u+1$ be $\delta_{u}(z)$. In other words, Algorithm \ref{alg:Pre-computation of delta} calculates the output message $\delta_{u}(z)$ corresponding to the input message $z$ from the perspective of the remaining graph. 

Consider a message update scheduling such that the message $x_{u}$ is updated only after the other messages are saturated to their fixed values. Under this message update scheduling, $x_{u}$ is updated from $z$ to $\epsilon \lambda_{u}(\delta_u(z))$ because the incoming message to the variable nodes at position $u$ is $\delta_{u}(z)$. In other words, pre-computation of $\delta_{u}(z)$ gives a one-dimensional DE equation $z^{(\ell+1)}=\epsilon \lambda_{u}(\delta_u(z^{(\ell)}))$ for position $u$ with an initial value of $z^{(0)}=\epsilon$ under the message update scheduling described before. The one-dimensional DE makes it possible to design the SC-LDPC ensemble by the methods used to optimize uncoupled LDPC codes, such as maximizing the design rate \cite{Richardson} and minimizing the number of required iteration \cite{Jamali}, \cite{Smith}. 

\subsection{Maximizing Design Rate}

\begin{algorithm}[h]
 \caption{Design method for maximizing design rate}
 \label{alg2}
 \begin{algorithmic}[1]
 \renewcommand{\algorithmicrequire}{\textbf{Input:} }
 \REQUIRE $l,r,L,w,l_{\min},l_{\max},Q,I_{\max}$
\STATE {\bf Initialization}: Code parameters of the regular SC-LDPC ensemble
	\FOR {${\rm Iter}=1$ to $I_{\max}$}
		\FOR {$u=L/2$ to $1$}
			\STATE Calculate the BP threshold $\epsilon^{\rm BP}$ by the DE equations (\ref{Eq:DE})
			\STATE Pre-computation of $\delta_{u}(z)$ by {\bf Algorithm \ref{alg:Pre-computation of delta}}
			\STATE Obtain $\lambda^{*}(x)=\sum\limits_{k=l_{\min}}^{l_{\max}}\lambda^{*}_kx^{k-1}$ by solving the following LP problem
			\begin{align*}
				&{\rm minimize~} \frac{1}{\int_{0}^{1}\lambda^{*}(x) dx}\\
				&{\rm subject~to~} \epsilon^{\rm BP}\lambda^{*}(\delta(z_q))<z_q {\rm~for~} 1\le q \le Q, \lambda^{*}(1) = 1
			\end{align*}
			\STATE  $\lambda_{u}(x)=\lambda^{*}(x),\lambda_{L-u+1}(x)=\lambda^{*}(x)$
		\ENDFOR
	\ENDFOR
 \end{algorithmic}
 \end{algorithm}
 
\begin{figure}[t]
\centering
\includegraphics[scale=0.39]{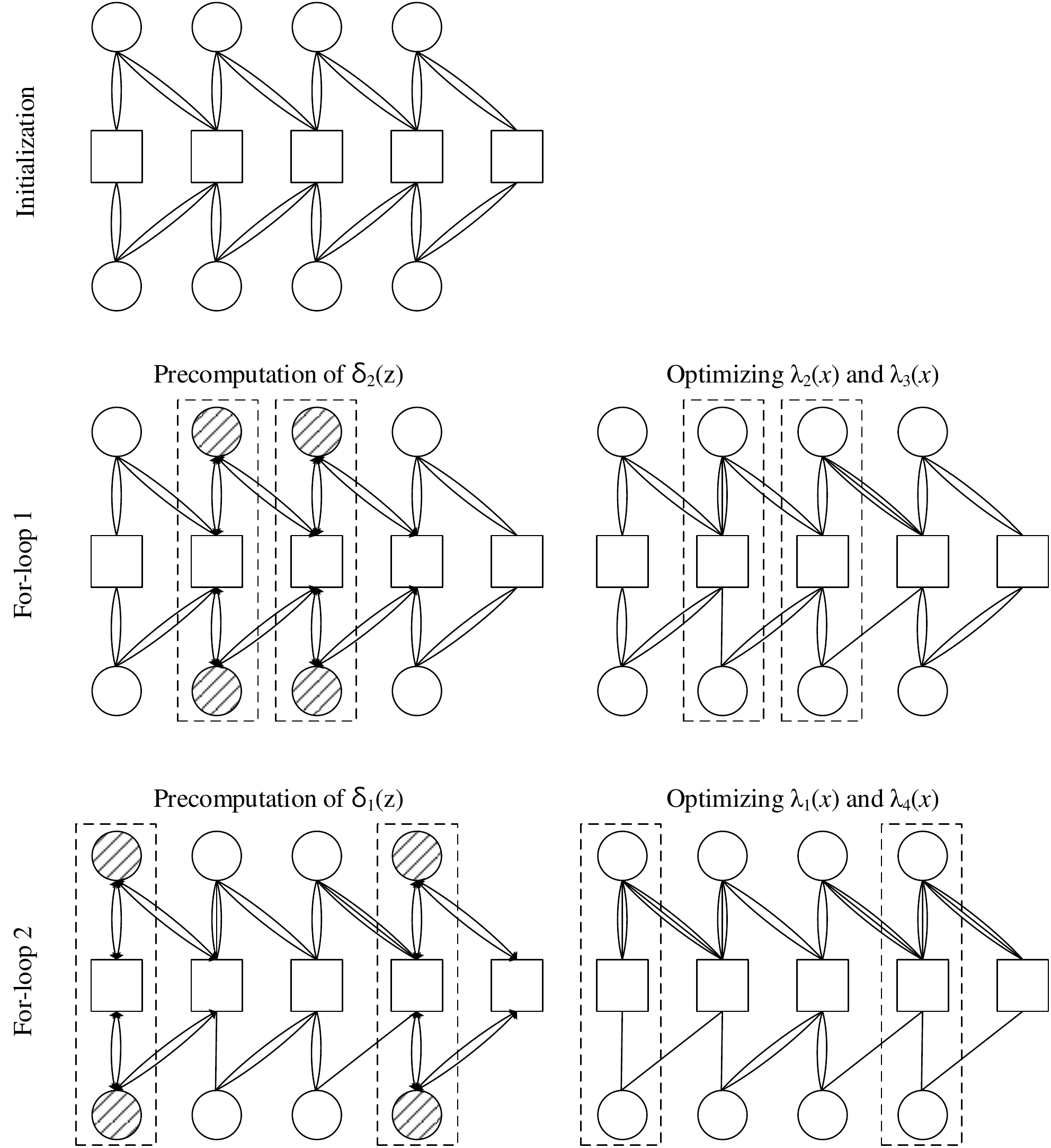}
\caption{Design procedure of the proposed algorithms with $L=4, w=2,$ and $M=2$.} 
\label{fig:opt}
\end{figure}

First, a design method to maximize the design rate is described in Algorithm \ref{alg2}. Maximizing the design rate can be achieved by solving the LP problem with the objective function to minimize the average variable node degree $1/{\int_{0}^{1}\lambda_u(x) dx}$ under the constraint of the successful decoding $\epsilon^{\rm BP}\lambda_u(\delta_u(z_q))<z_q {\rm~for~} q\in[1,Q]$. Unlike the codes defined in Section \ref{Sec:code_construction}, where the number of variable nodes at each position is fixed to $M$, there are $Ml{\int_{0}^{1}\lambda_{u}(x) dx}$ variable nodes at position $u$ because $Ml$ edges heading to the variable nodes at each position are distributed to the variable nodes with the average variable node degree $1/{\int_{0}^{1}\lambda_{u}(x) dx}$. Thus, minimizing the average degree $1/{\int_{0}^{1}\lambda_{u}(x) dx}$ corresponds to maximizing the number of variable nodes $Ml{\int_{0}^{1}\lambda_{u}(x) dx}$ and the design rate.

In Algorithm \ref{alg2}, the graph is initialized with the regular SC-LDPC ensemble. Then, the degree distribution $\lambda_{u}(x)$ is designed starting from $u=L/2\in\mathbb{N}$ to maximize the design rate and $\lambda_{L-u+1}(x)$ is also determined by the designed degree distribution because of the symmetric code structure. To design $\lambda_{L/2}(x)$, the value $\delta_{L/2} (z)$ is pre-calculated; this is used to design $\lambda_{L/2} (x)$ to maximize the design rate. The degree distribution for the next position $\lambda_{L/2-1}(x)$ is designed in the same manner after designing $\lambda_{L/2}(x)$. This local design is conducted until the number of iterations for the algorithm reaches $I_{\max}$ or $\lambda_{u}(x)$ does not change with $\rm{Iter}$. For the other algorithms to be introduced, we apply the same design procedure, in which the local design of the degree distributions is conducted for each pair of positions one at a time, as in Fig.~\ref{fig:opt}.

\begin{figure}[t]
\centering
\includegraphics[scale=0.30]{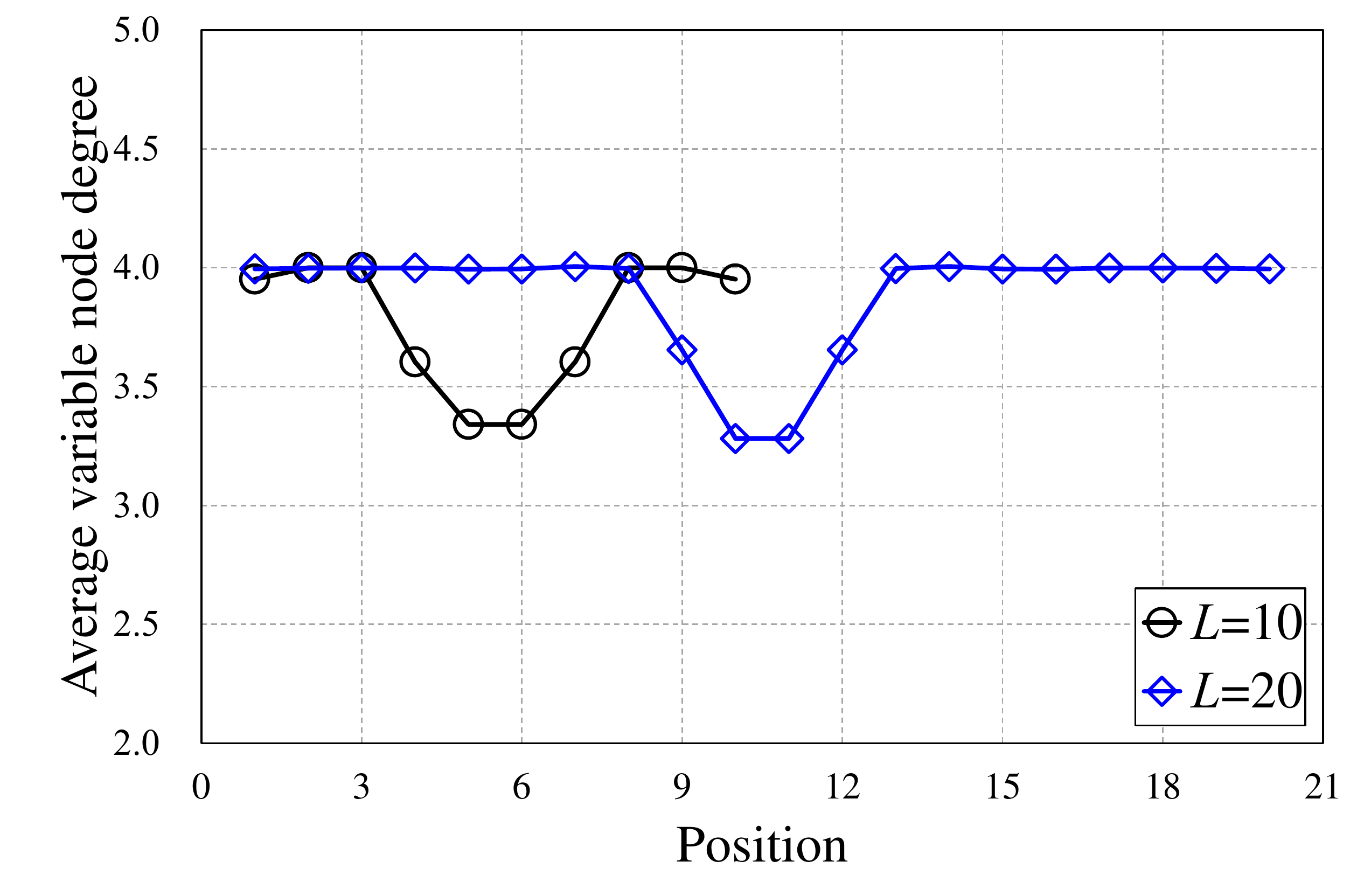}
\caption{Average variable node degree of each position for the SC-LDPC ensemble obtained by Algorithm 2.} 
\label{fig:Alg2}
\end{figure}

For the input values $l=4,r=8, L=10, w=3,l_{\min}=3, l_{\max}=10,Q=1000$, and $I_{max}=10$, the design rate is increased from $0.4$ to $0.4360$ by Algorithm \ref{alg2}. Likewise, the design rate is increased from $0.45$ to $0.4670$ by Algorithm \ref{alg2} for $L=20$. Fig.~\ref{fig:Alg2} shows the average variable node degree $1/{\int_{0}^{1}\lambda_{u}(x) dx}$ of each position for the SC-LDPC ensemble obtained by Algorithm \ref{alg2} for $L=10$ and $20$. It shows that the degree distributions of the center positions $5, 6$ for $L=10$ and $10, 11$ for $L=20$ are properly designed to increase the design rate by minimizing the corresponding average variable node degrees. However, the variable node degrees of the positions far from the center positions remain nearly identical to the initial degree $l=4$. Thus, the fraction of the number of positions to be improved by Algorithm \ref{alg2} becomes smaller as $L$ becomes larger. To achieve an improvement in all positions for large $L$, a different objective function for the LP problem is required instead of maximizing the design rate.

\subsection{Minimizing the Number of Required Iterations}
Another objective function described in Algorithm \ref{alg3} is minimizing the number of required iterations. Consider a non-increasing iterative function $f(x)$. Given an initial value $a$, the iterative process is represented by $x^{(0)}=a$ and $x^{(\ell)}=f(x^{(\ell-1)})$. The number of required iterations $I(b)$ for the target value $b$, $b<a$, is defined as the minimum $\ell$ such that $x^{(\ell)}\le b$. In \cite{Jamali}, the number of required iterations $I(b)$ is approximated as
\begin{align*}
I(b)\approx \sum\limits_{i=2}^{n-1}\frac{\Delta x_i}{x_i-f(x_i)}
\end{align*}
where $x_0=b, x_n=a, x_i=b+i\frac{a-b}{n}$, and $\Delta x_i=\frac{x_{i+1}-x_{i-1}}{2}$.

\begin{algorithm}[h]
 \caption{Design method for minimizing the number of required iterations}
 \label{alg3}
 \begin{algorithmic}[1]
 \renewcommand{\algorithmicrequire}{\textbf{Input:} }
 \REQUIRE $l,r,L,w,l_{\min},l_{\max},Q,I_{\max}$
\STATE {\bf Initialization}: Code parameters of the regular SC-LDPC ensemble
	\FOR {${\rm Iter}=1$ to $I_{\max}$}
		\FOR {$u=L/2$ to $1$}
			\STATE Calculate the BP threshold $\epsilon^{\rm BP}$ by the DE equations (\ref{Eq:DE})
			\STATE Pre-computation of $\delta_{u}(z)$ by {\bf Algorithm \ref{alg:Pre-computation of delta}}
			\STATE Obtain $\lambda^{*}(x)=\sum\limits_{k=l_{\min}}^{l_{\max}}\lambda^{*}_kx^{k-1}$ by solving the following LP problem
			\begin{align*}
				&{\rm minimize~} \sum\limits_{q=2}^{Q-1}\frac{\Delta z_q}{z_q-\epsilon^{\rm BP}\lambda_u(\delta_u(z_q))}\\
				&{\rm subject~to~} \frac{1}{\int_{0}^{1}\lambda^{*}(x) dx}=l \\
								&\epsilon^{\rm BP}\lambda^{*}(\delta_u(z_q))<z_q {\rm~for~} 1\le q \le Q, \lambda^{*}(1) = 1
			\end{align*}
			\STATE  $\lambda_{u}(x)=\lambda^{*}(x),\lambda_{L-u+1}(x)=\lambda^{*}(x)$
		\ENDFOR
	\ENDFOR
 \end{algorithmic}
 \end{algorithm}

\begin{figure}[t]
\centering
\includegraphics[scale=0.30]{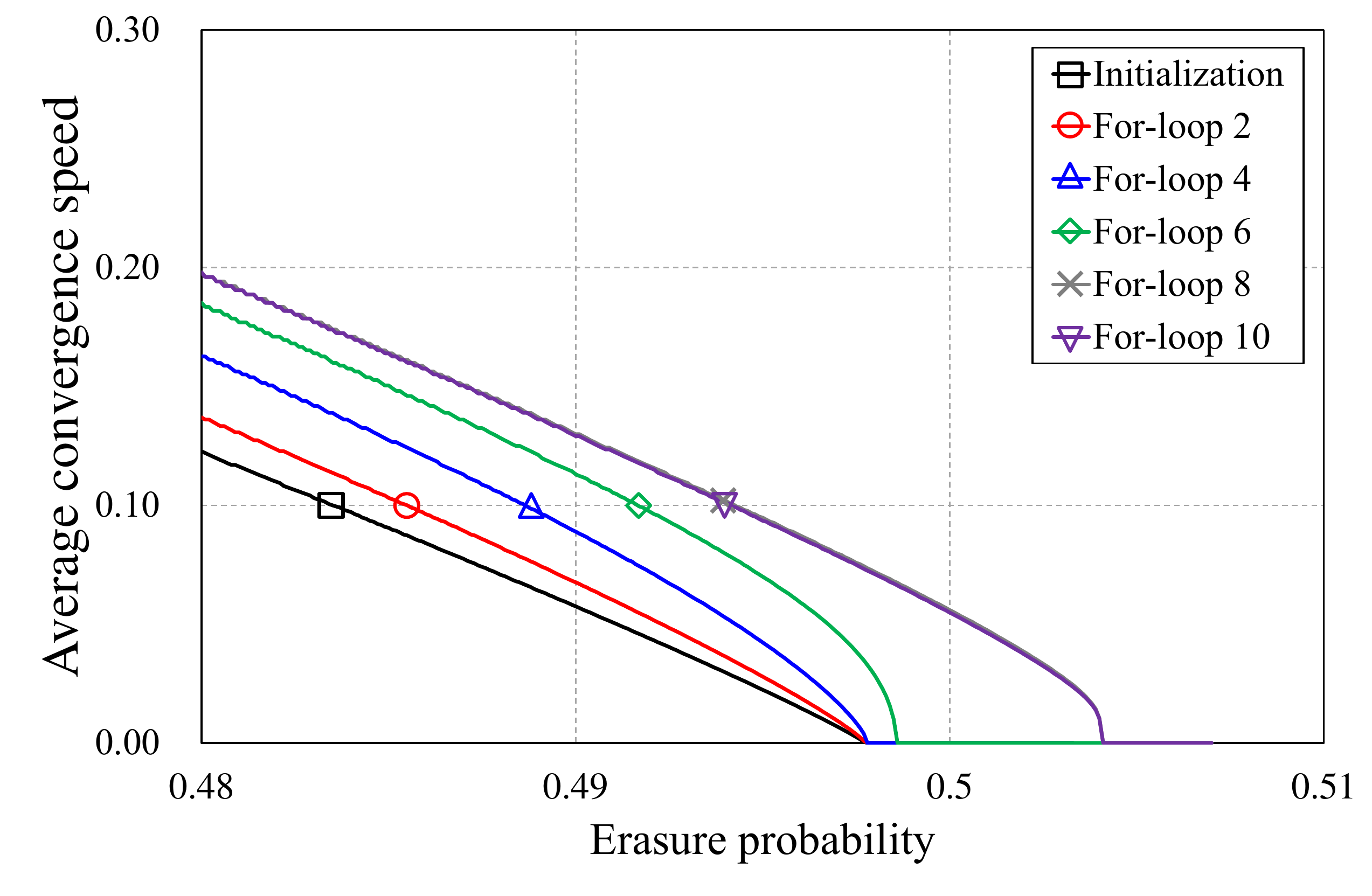}
\caption{Evolution of average convergence speed as the for-loop in Algorithm \ref{alg3} proceeds.} 
\label{fig:alg3}
\end{figure}


\begin{table}[t]
\centering
\caption{Coefficients of $\lambda_{u}(x)$ for the SC-LDPC ensemble obtained by Algorithm 3 for $L=20$}
\begin{tabular}{c|cccccccc}
\hline
$u$ & $x^2$    & $x^3$    & $x^4$    & $x^5$    & $x^6$    & $x^7$    & $x^8$    & $x^9$    \\ \hline
1   & 0        & 1        & 0        & 0        & 0        & 0        & 0        & 0        \\
2   & 0.1916 & 0.4889 & 0.3194 & 0        & 0        & 0        & 0        & 0        \\
3   & 0.3852 & 0        & 0.5769 & 0.0362 & 0        & 0        & 0.0017 & 0        \\
\vdots   &  &         &  & \vdots &        &         &         &         \\
7   & 0.6128 & 0        & 0        & 0        & 0        & 0.2644 & 0.0357 & 0.0870 \\
8   & 0.6429 & 0        & 0        & 0        & 0        & 0        & 0        & 0.3571 \\
9   & 0.6429 & 0        & 0        & 0        & 0        & 0        & 0        & 0.3571 \\
10  & 0.6429 & 0        & 0        & 0        & 0        & 0        & 0        & 0.3571 \\ \hline
\end{tabular}\label{Table:DDs_Alg3}
\end{table}

In the case of the one-dimensional DE $z^{(\ell+1)}=\epsilon \lambda_{u}(\delta_u(z^{(\ell)}))$ of the SC-LDPC ensemble, the iterative function is given as $f(x)=\epsilon \lambda_{u}(\delta_u(x))$. Thus, the number of required iterations from the initial value $z_Q=\epsilon$ to the target value $z_1=\epsilon/Q$ is approximated as
\begin{equation}
 \sum\limits_{q=2}^{Q-1}\frac{\Delta z_q}{z_q-\epsilon\lambda_u(\delta_u(z_q))}
 \label{Eq:Minmizing iteration}
\end{equation}
where $\Delta z_q=1/Q$. Thus, the design method that minimizes the number of iterations can be summarized as Algorithm \ref{alg3}, where the objective function is to minimize the equation (\ref{Eq:Minmizing iteration}). In Algorithm \ref{alg3}, there is the additional constraint ${1}/{\int_{0}^{1}\lambda^{*}(x) dx}=l $, which implies that the average variable node degree should be equal to the variable node degree $l$ of the initial regular SC-LDPC ensemble to maintain the design rate. As an example, some part of coefficients of $\lambda_{u}(x)$ obtained by Algorithm \ref{alg3} with $l=4,r=8, L=20, w=3,l_{\min}=3, l_{\max}=10,Q=1000$, and $I_{max}=10$ are shown in Table \ref{Table:DDs_Alg3}. Note that, to maintain the advantages of the regular SC-LDPC codes whose minimum variable node degree is larger than $2$ \cite{Costello}, we generally initialize the algorithm from the regular SC-LDPC ensemble with $l=4$ and the minimum degree constraint as $l_{\min}=3$. Instead, Algorithm \ref{alg3} is not initialized from the regular SC-LDPC ensemble with $l=3$ because a non-zero fraction of degree-two variable nodes is unavoidably required to introduce a fraction of variable nodes with degree larger than $3$. 

 In order to investigate the change of the overall number of required iterations in the DE equations (\ref{Eq:DE}) at each local design, the evolution of the average convergence speed is shown in Fig.~\ref{fig:alg3} as the for-loop in Algorithm \ref{alg3} proceeds. It shows that the average convergence speed increases steadily as the local design of each position proceeds. In addition, we note an increase in the BP threshold at for-loop 6, where the BP threshold corresponds to the erasure probability that the average convergence speed becomes zero.

\begin{figure*}[t]
\centering
\subfigure[Algorithm 2]{\includegraphics[scale=0.30]{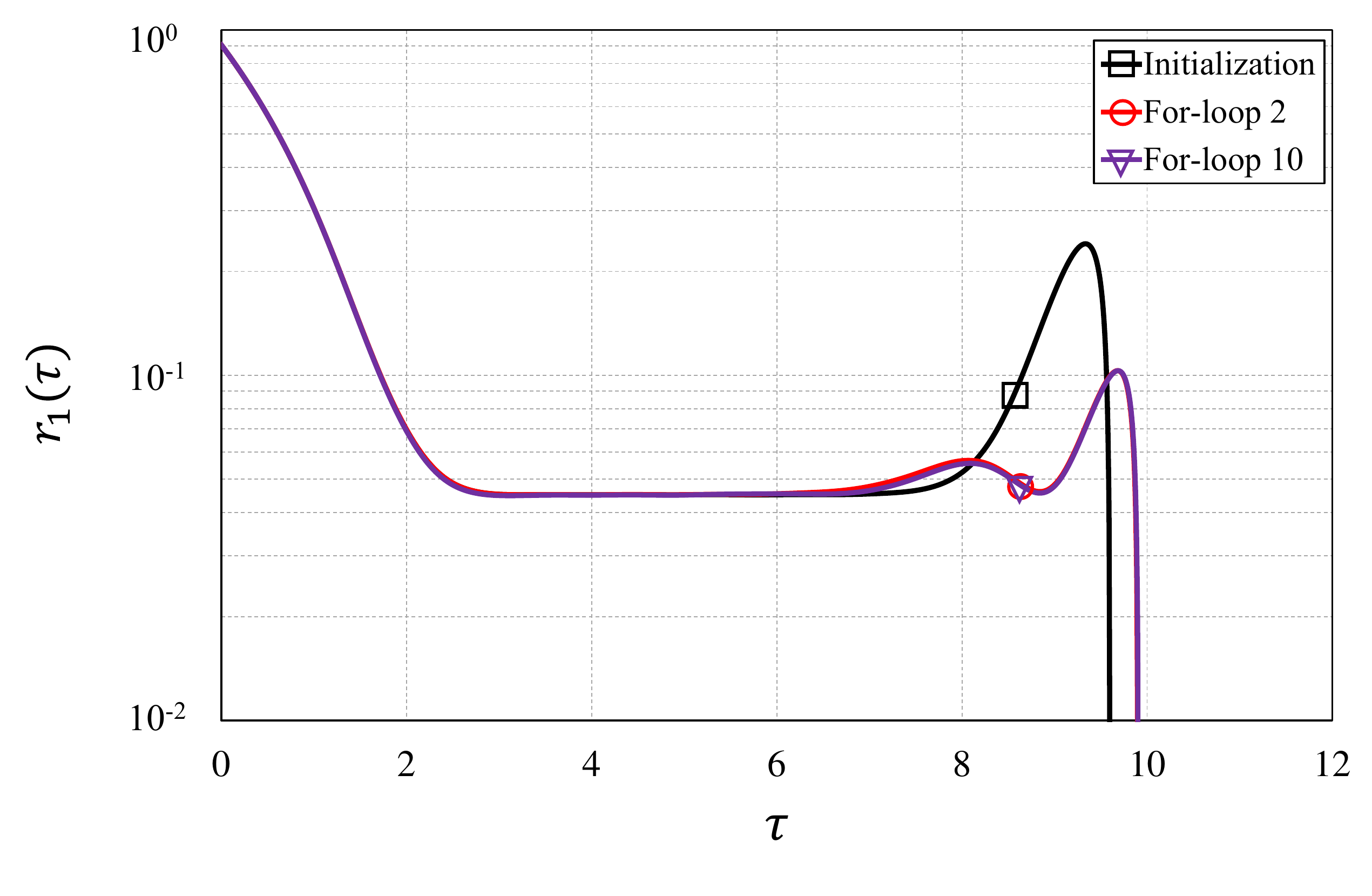}}
\subfigure[Algorithm 3]{\includegraphics[scale=0.30]{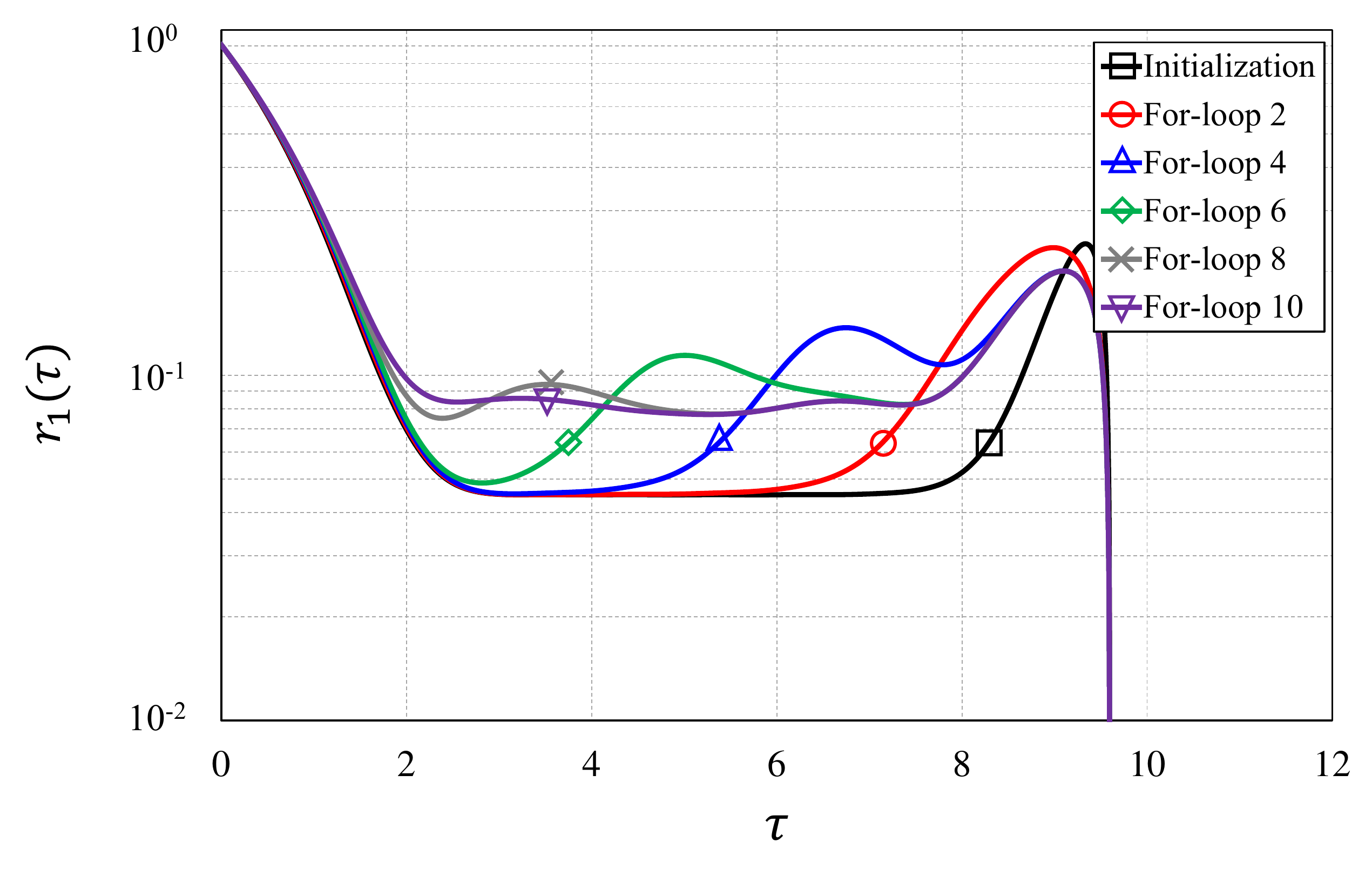}}%
\caption{Evolution of ${r}_{1}(\tau)$ for the SC-LDPC ensembles obtained by Algorithms \ref{alg2} and \ref{alg3} as the for-loop proceeds.}
\label{Fig:EGE}
\end{figure*}

To observe the change in the SC-LDPC ensemble according to the local design of Algorithms \ref{alg2} and \ref{alg3} graphically, the expected graph evolution can be used. With the differential equations in Section \ref{Sec:EGE}, Fig.~\ref{Fig:EGE} shows the evolutions of ${r}_{1}(\tau)$ at $\epsilon=0.48$ for the SC-LDPC ensembles obtained by the algorithms as the for-loop in the algorithms proceeds. The evolution of $r_1(\tau)$ at the initialization corresponds to $r_1(\tau)$ for the regular SC-LDPC ensemble that is analyzed in \cite{Olmos}. According to the analysis in \cite{Olmos}, $r_1(\tau)$ remains constant at the local minimum for a certain interval of $\tau$, referred to as the critical phase. On the basis of the critical phase, the evolution of $r_1(\tau)$ is divided into three phases: The initial phase, the critical phase, and the third phase. Most of the variable nodes located in the boundary positions have already been recovered in the initial phase. Then, in the critical phase, two decoding waves emerge and travel along the inner positions while recovering the variable nodes through which the decoding waves pass. If the local performance of each position is defined as the number of degree-one check nodes when the decoding waves pass through the position, the steady value of $r_1(\tau)$ in the critical phase implies that the local performances of the inner positions are all the same. Moreover, it can be said that the BP threshold of the SC-LDPC ensemble is determined by the local performance of the inner positions. Finally, in the third phase, the decoding waves meet around the center and the variable nodes located at the center positions are recovered, which corresponds to the increase in the $r_1(\tau)$ value. 

Fig.~\ref{Fig:EGE}(a) shows the evolution of $r_1(\tau)$ as the for-loop in Algorithm \ref{alg2} proceeds. After performing for-loop 2 in Algorithm \ref{alg2}, the degree distributions of the center positions from $9$ to $12$ are locally designed to decrease their average variable node degree. However, the value of $r_1(\tau)$ is decreased in the third phase at the expense of lowering the average variable node degrees of the center positions as in Fig.~\ref{Fig:EGE}(a). In addition, it is observed that the minimum value of $r_1(\tau)$ in the third phase is nearly identical to the value of $r_1(\tau)$ in the critical phase due to the constraint in Algorithm \ref{alg2} that the BP threshold should be maintained. If the minimum value of $r_1(\tau)$ in the third phase becomes smaller than the value of the critical phase, the BP threshold will be degraded. In other words, because the value of $r_1(\tau)$ in the third phase is larger than the critical phase value for the initial regular SC-LDPC ensemble, there is a room for improvement in the design rate for the center positions. However, an improvement in the design rate of the inner positions cannot be achieved while maintaining the BP threshold. Accordingly, the average variable node degree of the positions except the center positions does not change, as indicated in Fig.~\ref{fig:Alg2}, and the evolution of $r_1(\tau)$ is unchanged from for-loops $2$ to $10$.

Compared to Algorithm \ref{alg2}, Fig.~\ref{Fig:EGE}(b) shows the evolution of $r_1(\tau)$ as the for-loop in Algorithm \ref{alg3} proceeds. After for-loop 2, where the local design has been performed from center positions $9$--$12$, $r_1(\tau)$ in the right part of the critical phase increases. This means that the local design improves the local performance of the center positions and accordingly the length of the critical phase is shortened and the overall number of required iterations is decreased. Likewise, from for-loops 2 to 4, the length of the critical phase is steadily shortened. After for-loop 6, the critical phase observed in the initial regular SC-LDPC ensemble disappears and only a single local minimum point remains. The local minimum value is greater than the value of $r_1(\tau)$ in the critical phase, which means that the BP threshold is increased. Further, in for-loop 8, the critical phase observed in the initialization completely disappears, which implies that the BP threshold is increased significantly. In other words, the BP threshold is increased in for-loop 6 because the local performances of positions $7$--$14$ have already been improved. Conversely, the BP threshold does not increase up to for-loop $4$ because the local performances of positions $1$--$6$ and $15$--$20$, which are not yet designed at this point, become a bottleneck despite the fact that the local performances of positions $7$--$14$ are improved by the local design. On the other hand, the BP threshold is increased when the bottleneck positions $5$ and $16$ are designed in for-loop $6$. In summary, the BP threshold can be increased if the degree distributions of all positions are properly designed by Algorithm \ref{alg3}. In Table \ref{table:BP}, the BP thresholds of the regular SC-LDPC ensemble and the SC-LDPC ensembles obtained by Algorithm \ref{alg3} are shown. 

\begin{table}[t]
\centering
\caption{Comparing the BP thresholds of the regular SC-LDPC with the ensembles obtained by the proposed algorithms}
\label{table:BP}
\begin{tabular}{l|ccc}
\hline
                               & \multicolumn{3}{c}{$\epsilon^{\rm BP}$} \\
                               & $L=10$       & $L=20$      & $L=30$      \\ \hline
SC-LDPC, Regular               & $0.4981$     & $0.4977$    & $0.4977$    \\
SC-LDPC, Alg. 3                & $0.5241$     & $0.5069$    & $0.5027$    \\
SC-LDPC, Alg. 4 		 & $0.5679$     & $0.5247$    & $0.5087$    \\ \hline
\end{tabular}
\end{table}

\begin{figure}[t]
\centering
\includegraphics[scale=0.30]{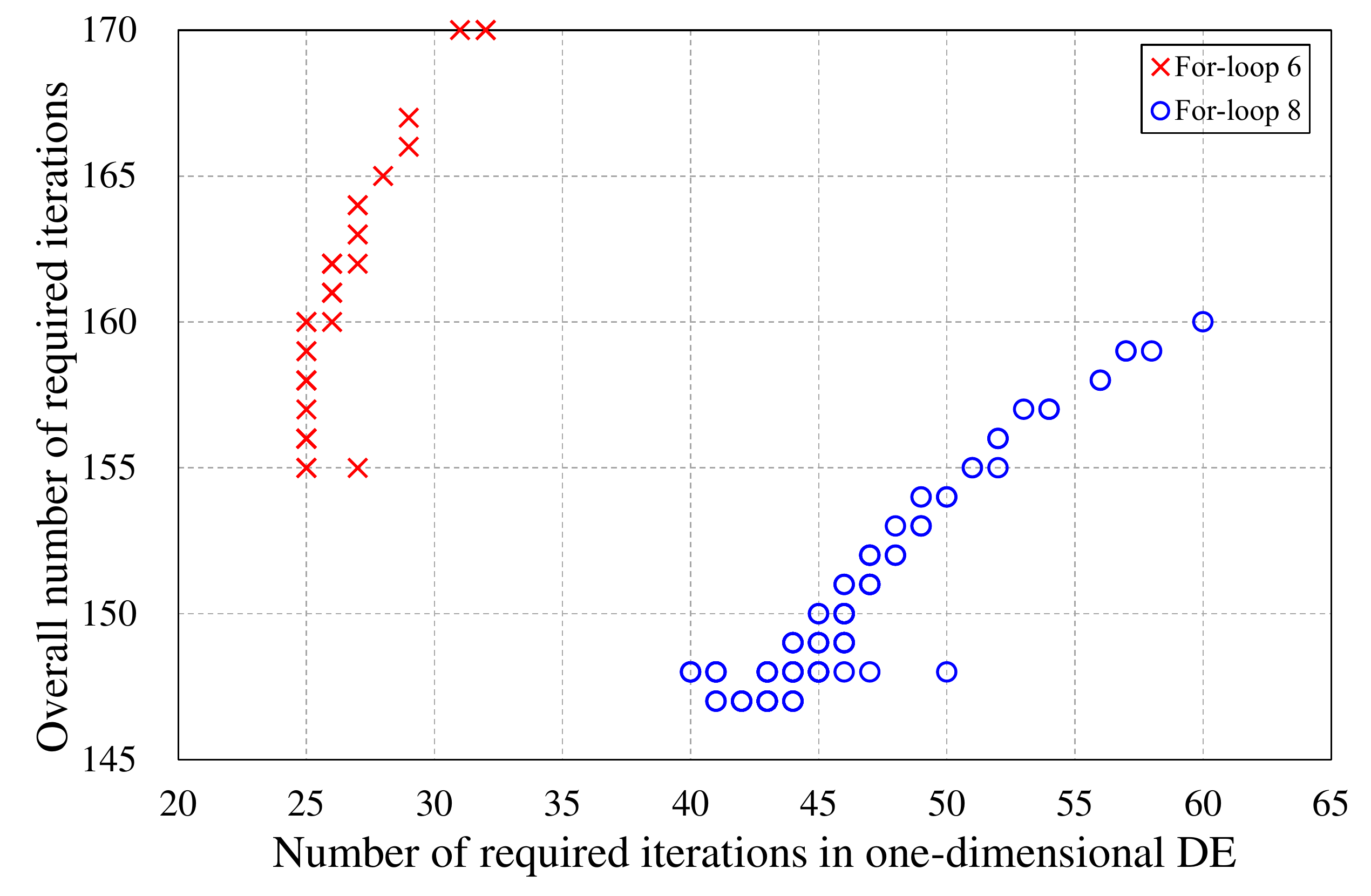}
\caption{Correlation between the two kinds of numbers of required iterations.} 
\label{fig:Iteration}
\end{figure}

It is important to remark that the degree distributions obtained by Algorithm \ref{alg3} are not the global optimum that minimizes the overall number of required iterations in the DE equations (\ref{Eq:DE}) because of the two facts. First, the local design is performed instead of the global design. Second, the one-dimensional DE assumes different decoding scheduling from the conventional parallel scheduling as mentioned in Section~\ref{Sec:pre-computation}. Therefore, minimizing the number of required iterations in the one-dimensional DE equation does not guarantee minimization of the overall number of required iterations in parallel scheduling. However, we numerically observe that these two kinds of numbers of required iterations are generally proportional. Fig.~\ref{fig:Iteration} shows the correlation between the numbers of required iterations in the one-dimensional DE equation $z^{(\ell+1)}=\epsilon \lambda_{u}(\delta_u(z^{(\ell)}))$ and in DE equations (\ref{Eq:DE}) at for-loops 6 and 8 in Algorithm \ref{alg3}. Each point in Fig.~\ref{fig:Iteration} is the numbers of required iterations measured at $\epsilon=\epsilon^{\rm BP}-0.01$ for a randomly generated degree distribution. According to Fig.~\ref{fig:Iteration}, it can be seen that both numbers of required iterations are generally proportional and thus the objective function of Algorithm \ref{alg3} is proper for reducing the overall number of required iterations.

\subsection{Minimizing the Number of Required Iterations with Non-Uniform Check Node Degrees}

\begin{algorithm}[h]
 \caption{Design method for minimizing the number of required iterations for the SC-LDPC ensemble with non-uniform check node degrees}
 \label{alg4}
 \begin{algorithmic}[1]
 \renewcommand{\algorithmicrequire}{\textbf{Input:} }
 \REQUIRE $l,r,L,w,M,l_{\min},l_{\max},r_{\min},r_{\max},Q,I_{\max}$
\STATE {\bf Initialization}: Code parameters of the regular SC-LDPC ensemble
	\FOR {${\rm Iter}=1$ to $I_{\max}$}
		\FOR {$u=L/2$ to $1$}
		\STATE Calculate the BP threshold $\epsilon^{\rm BP}$ by the DE equations (\ref{Eq:DE}) and set $s=0$
			\FOR {$v=u$ to $u+w-1$}
				\FOR {$r_v=r_{\min}$ to $r_{\max}$}
				\STATE  ${T}_{v,u}=r_{v}M\frac{l}{r}-\sum\limits_{j\neq u}{T}_{v,j}$, $r_{L+w-v}=r_{v}$, ${T}_{L+w-v,L-u+1}={T}_{v,u}$
				\STATE $s=s+1$, $v^s=v, r^s=r_v, {\vect T}^s={\vect T}$
				
				\STATE Pre-computation of $\delta_{u}(z)$ by {\bf Algorithm \ref{alg:Pre-computation of delta}}
				\STATE Obtain $\lambda^{*}_{s}(x)=\sum\limits_{k=l_{\min}}^{l_{\max}}\lambda^{*}_{s,k}x^{k-1}$ by solving the following LP problem
				\begin{align*}
					&{\rm minimize~}I(s)=\sum\limits_{q=2}^{Q-1}\frac{\Delta z_q}{z_q-\epsilon^{\rm BP}\lambda^{*}_{s}(\delta_u(z_q))} \\
					&{\rm subject~to~} \frac{1}{\int_{0}^{1}\lambda^{*}_{s}(x) dx}={\sum_{i}{T}_{i,u}}/{M},\\ 
					&\epsilon^{\rm BP}\lambda_{s}^{*}(\delta_u(z_q))<z_q {\rm~for~} 1\le q \le Q, \lambda_{s}^{*}(1) = 1
				\end{align*}
					\ENDFOR
				\ENDFOR
			\STATE $s^{\rm *}=\argmin I(s)$
			\STATE $\lambda_{u}(x)=\lambda_{s^{*}}^{*}(x),\lambda_{L-u+1}(x)=\lambda^{*}_{s^{*}}(x), r_{v}=r^{s^{*}},r_{L+w-v}=r^{s^{*}}, {\vect T}={\vect T}^{s^*}$
		\ENDFOR
	\ENDFOR
 \end{algorithmic}
 \end{algorithm}

In Algorithm \ref{alg3}, we consider the SC-LDPC ensemble with a uniform check node degree. However, additional improvements can be achieved by permitting non-uniform check node degrees for the SC-LDPC ensemble, which is considered in Algorithm \ref{alg4}. For example, while designing $\lambda_{u}(x)$, the degree $r_u$ is changed from $r_{\min}$ to $r_{\max}$ and the optimal $r_u$ value is selected such that the approximated value of the number of required iterations in (\ref{Eq:Minmizing iteration}) is minimized. Such comparison is performed for all degrees of check nodes connected to the variable nodes at position $u$, that is, $r_u,\ldots,r_{u+w-1}$ and then the optimal result is obtained. Note that the corresponding entries $T_{u,u},\ldots,T_{u+w-1,u}$ of the connectivity matrix ${\vect T}$ should be modified according to the selected check node degree. This algorithm permits non-uniform check node degrees for each position and a different connectivity matrix from that of the regular SC-LDPC ensemble. Because it adds a degree of freedom in the design of $\lambda_{u}(x)$, Algorithm \ref{alg4} has higher complexity than Algorithm \ref{alg3} but gives code parameters.

Unlike Algorithm \ref{alg3}, Algorithm \ref{alg4} can be initialized with the regular SC-LDPC ensemble with $l=3$ and the minimum variable node degree $l_{\min}=3$ because the average variable node degree ${1}/{\int_{0}^{1}\lambda_{u}(x) dx}$ can be higher than $l=3$ if the degrees of the connected check nodes are increased. For example, consider when the positions $u=L/2$ and $L/2+1$ are firstly designed for $(l,r)=(3,6)$ and $w=3$. Recall that the number of edges $\sum_i T_{i,u}$ coming into the variable nodes at position $u$ should be equal to the number of sockets ${M}/{\int_{0}^{1}\lambda_{u}(x) dx}$ and that there are ${Ml}/{w}=M$ edges between two connected positions. Thus, the average variable node degree ${1}/{\int_{0}^{1}\lambda_{u}(x) dx}$ of the variable nodes at position $u$ should be equal to $3$ because the number of edges coming to the variable nodes at position $u$ is $\sum_i T_{i,u}=3M$ for the regular SC-LDPC ensemble. However, if the degree of check nodes at position $u$ is increased from $6$ to $7$ and the number of edges between variable nodes and check nodes at position $u$ is accordingly increased from $M$ to $\frac{3}{2}M$ as an example, the number of edges connected to variable nodes at position $u$ becomes $\frac{7}{2}M$. Then the average variable node degree becomes $\frac{7}{2}\ge 3$. Therefore, variable nodes of higher degrees can be introduced while maintaining the minimum variable node degree $l_{\min}$ as $3$.

Table \ref{table:BP} summarizes the BP thresholds of the SC-LDPC ensembles obtained by the proposed algorithms. Since the BP thresholds are better with $(l,r)=(3,6)$ than with $(l,r)=(4,8)$ for the ensembles obtained by Algorithm \ref{alg4}, the BP thresholds of the ensembles obtained by Algorithm \ref{alg4} with $(l,r)=(3,6)$ are included in Table \ref{table:BP}.
The connectivity matrix ${\vect T}$ for the degree distributions obtained by Algorithm \ref{alg4} with $l=3,r=6, L=20, w=3,l_{\min}=3, l_{\max}=10, r_{\min}=6, r_{\max}=8, Q=1000$, $I_{\max}=10$, and $M=6$ is
\begin{equation*}
\arraycolsep=2pt\def\arraystretch{0.7}{\vect T}=\left[\small\begin{array}{cccccc}
12 &    &      &      &      &       \\
3 & 15  &      &      &      &       \\
6 & 6  & 12    &      &      &      \\
  & 9 & 3   & 12    &      &       \\
  &    & 9   & 3    & 12    &       \\
  &    &      & 12   & 6    & \ddots   \\
  &    &      &      &6    & \ddots  \\
  &    &      &      &      & \ddots  \\
\end{array}\right].
  \end{equation*}
As you can see, the non-zero components of the connectivity matrix are not constant. The resulting Tanner graph corresponds to the case of Graph 3 in Fig.~\ref{Fig:Example}.

\section{Finite-Length Code Performances}
In this section, we introduce several methods which can be used to improve their performance further and show validity of the proposed design algorithms for SC-LDPC ensembles by comparing the finite-length code performance. 



\subsection{Performance Improvement by Multi-Edge Type Check Nodes}\label{Subsec:MET}
The SC-LDPC ensemble defined in Section \ref{Sec:code_construction} is referred to as the randomly constructed ensemble because the variable node sockets at position $u$ are connected to the check nodes at position $v$ at random with probability ${T}_{v,u}/\sum_{i}{ T}_{i,u}$. However, as mentioned in previous works \cite{Mitchell}, \cite{Stinner}, \cite{Olmos}, randomly constructed SC-LDPC codes are inferior to SC-LDPC codes with specific structures such as protograph-based SC-LDPC codes in terms of the finite-length code performance. Protograph-based SC-LDPC codes have the MET structure \cite{Richardson} for both variable and check nodes. However, in this paper, we consider the SC-LDPC ensemble having the MET structure only for check nodes in designing the degree distributions because the edges of variable nodes should be connected randomly to utilize the one-dimensional DE. 


It is assumed that the edges connected to a check node consist of $w$ edge types. An edge of type $t$, which is connected to a check node at position $v$, is connected to a variable node at position $v-t+1$ for $1\le t \le w$. A degree type of check nodes is represented by the vector $\underline{d}=({d}_1,\ldots,{d}_{w})$ such that ${d}_t$ is the number of edges of type $t$. Also, at position $v$, it is assumed that there are $w$ degree types represented by each row of a $w\times w$ degree type matrix ${\vect S}^v$. To be specific, $M_c\frac{1}{w}$ check nodes have a degree type represented by $\underline{d}=\underline{S}_k^v\triangleq(S_{k,1}^v,\ldots,S_{k,w}^v)$, where $\underline{S}_k^v$ denotes the $k$th row vector of ${\vect S}^v$ for $k\in[1,w]$. For example, the MET regular SC-LDPC ensemble with $r=6$ and $w=3$ has $S_{k,j}^v=2$ for all $k$ and $j$,
 which means that all check nodes have the same degree type vector $(2,2,2)$ similar to the structure of the protograph-based SC-LDPC codes \cite{Mitchell}. For the regular SC-LDPC ensemble when $r/w$ is not an integer, degree types of check nodes are not yet defined in the protograph-based ensemble. To define these degree types as closely as possible to the structure in the protograph-based ensemble, the first row $\underline{S}_1^v$ of the degree type matrix ${\vect S}^v$ for position $v$ is defined as
\begin{equation*}
\underline{S}_1^v=\big(\overset{r {\rm~mod~}w}{\overbrace{\lceil{r/w}\rceil,\ldots,\lceil{r/w}\rceil},}\overset{w-(r {\rm~mod~}w)}{\overbrace{\lfloor{r/w}\rfloor,\ldots,\lfloor{r/w}\rfloor}}\big).
\end{equation*}
In addition, the $k$th row $\underline{S}_k^v$ for $2\le k \le w$ is obtained by right circular shifting the first row $\underline{S}_1^v$ by $k$ times. 

 The number of edges between the check nodes at position $v$ and the variable nodes at position $v-t+1$ becomes $M_c\frac{1}{w}\sum_{k}{S}^v_{k,t}$, which corresponds to ${T}_{v,v-t+1}$.
 Generally, the construction method of MET SC-LDPC codes is described as follows. First, the placement of variable and check nodes is identical to that of the randomly constructed SC-LDPC codes in Section \ref{Sec:code_construction}. We label the variable and check node sockets at each position and assign the degree type vector $\underline{S}_k^v$ to some $M_c \frac{1}{w}$ check nodes at position $v$ for $k\in[1,w]$. Then, the number of check node sockets connected to variable nodes at position $v-t+1$ is $M_c\frac{1}{w}\sum_{k}{S}^v_{k,t}={T}_{v,v-t+1}$ and let these check node sockets be group $t$ for $t\in [1,w]$. 
 Let $\pi_{u}$ be a random permutation on $[1,\sum_{i}{T}_{i,u}]$ for position $u$. Divide $\pi_{u}$ into $w$ disjoint subsets denoted by $\pi_{u}^{1},\ldots,\pi_{u}^{w}$ such that the size of $\pi_{u}^{t}$ becomes ${ T}_{u+t-1,u}$. Then, the $j$th check node socket in group $t$ at position $u+t-1$ is connected to the $\pi_{v}^{t}(j)$th variable node socket at position $u$.

Because the ${1}/{w}$ fraction of the check nodes at position $v$ have check node degree type $\underline{S}_k^v$, the node perspective degree distribution \cite{Richardson} of the check nodes at position $v$ is represented as
\begin{equation*}
R_{v}(\underline{x})=\sum\limits_{k=1}^{w}\frac{1}{w}\underline{x}^{\underline{S}_k^v}
\end{equation*}
where $\underline{x}=(x_1,\ldots,x_w)$ and $\underline{x}^{\underline{S}_k^v}$ denotes $\Pi_{t=1}^{w}x_t^{{ S}^{v}_{k,t}}.$
From the node perspective degree distribution, the edge perspective degree distribution of the check nodes at position $v$ is obtained as
\begin{align*}
\underline{\rho}_{v}(\underline{x})&=\left(\frac{\partial R_{v}(\underline{x})}{\partial x_1}/\frac{\partial R_{v}(\underline{1})}{\partial x_1},\ldots,\frac{\partial R_{v}(\underline{x})}{\partial x_w}/\frac{\partial R_{v}(\underline{1})}{\partial x_w}\right)\\
&=\left(\sum\limits_{k=1}^{w}\frac{{ S}^{v}_{k,1}}{r_v}{\underline x}^{{\underline S}^{v}_{k}}/x_{1},\ldots,\sum\limits_{k=1}^{w}\frac{{ S}^{v}_{k,w}}{r_v}{\underline x}^{{\underline S}^{v}_{k}}/x_{w} \right).
\end{align*}
Similar to the DE equations in (\ref{Eq:DE}), let $x_u^{(\ell)}$ denote the average erasure probability of messages at iteration $\ell$ emitted from the variable nodes at position $u$. In addition, let the $t$th element of a vector ${\underline y}_{v}^{(\ell)}=(y_{v,1}^{(\ell)},\ldots,y_{v,w}^{(\ell)})$ be the average erasure probability of messages at iteration $\ell$ emitted from the check nodes at position $v$ to the variable nodes at position $v-t+1$. Set the initial conditions as $x_u^{(0)}=\epsilon$ for all $u$. For simple expression of the DE equations, it is assumed that there exist variable nodes at position $u$ for $u<0$ with the assumption $x_u^{(\ell)}=0$ for all ${\ell}$. Then, the DE equations for the MET SC-LDPC ensembles are described as
\begin{align}
{\underline y}_{v}^{(\ell)}&=1-{\underline \rho}_{v}(1-x_{v-(w-1)}^{(\ell)},\ldots,1-x_{v}^{(\ell)})\nonumber\\
x_{u}^{(\ell+1)}&=\epsilon\lambda_{u}\left(\frac{\sum\limits_{t=1}^{w}{T}_{u+t-1,u}{y}_{u+t-1,t}^{(\ell)}}{\sum_{i}{ T}_{i,u}}\right).\label{Eq:DE_ME}
\end{align}

Algorithms 3 and 4 can be applied to the MET SC-LDPC ensemble using the DE equations in (\ref{Eq:DE_ME}). Note that especially for Algorithm $4$, the degree type matrix ${\bf S}^v$ should be changed according to changes of check node degrees and the connectivity matrix. Also, because the DE equations are changed as in (\ref{Eq:DE_ME}), the resulting degree distributions differ from those obtained by the randomly constructed SC-LDPC ensemble. 


\subsection{Applying the Proposed Algorithms to SC-RA Codes}

\begin{figure}[t]
\centering
\includegraphics[scale=0.45]{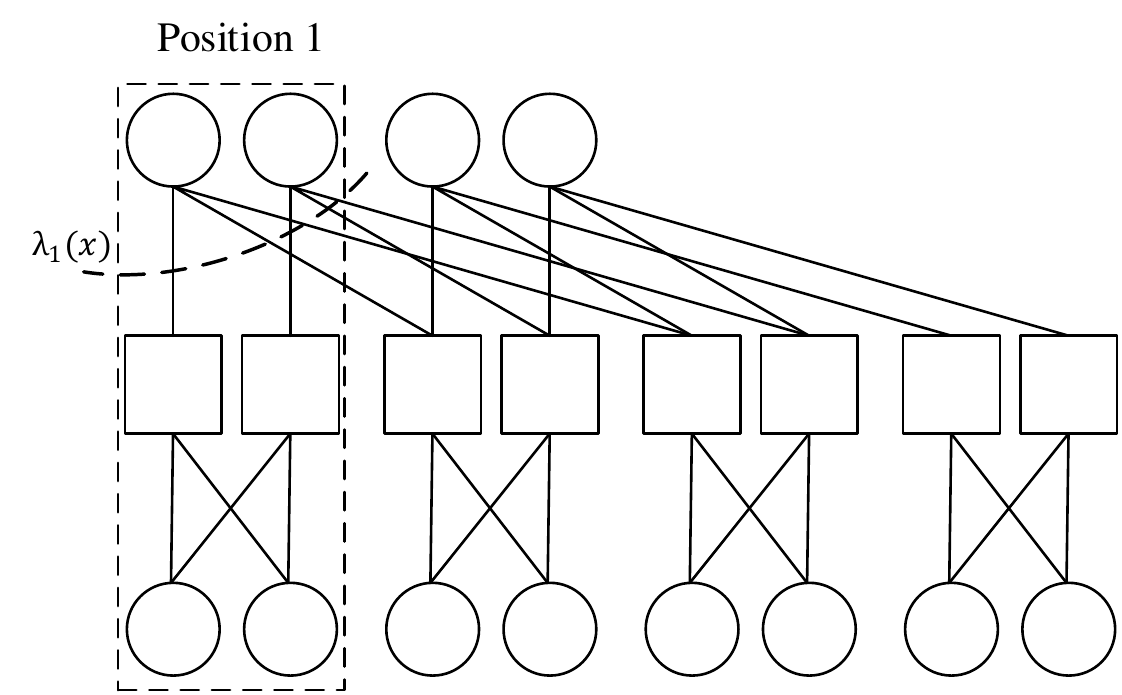}
\caption{A Tanner graph of the SC-RA ensemble for $q=3$, $L=2$, and $M=2$.} 
\label{fig:Graph_SC_RA}
\end{figure}

In this subsection, it is shown that the proposed algorithms can be applied to other coupled code structures. Especially, the SC-RA code ensemble \cite{Johnson} is considered due to its superior decoding performance reported in \cite{Stinner}. It is known that SC-RA codes outperform regular SC-LDPC codes with the same design rate, code length, and decoding complexity \cite{Stinner}, \cite{Johnson}. The SC-RA ensemble is constructed by coupling the $L$ disjoint uncoupled $(q,a)$-regular RA ensemble. Specifically, the SC-RA ensemble with $q=a$ is considered in this paper. Fig.~\ref{fig:Graph_SC_RA} shows a Tanner graph of the SC-RA ensemble for $L=2$, $q=3$, and $M=4$. There are $M/2$ variable nodes of degree $q$ at each position from $1$ to $L$ in the upper side and $M/2$ variable nodes of degree $2$ at each position from $1$ to $L+q-1$ in the lower side. The variable nodes of degree $q$ at position $u$ are connected to the check nodes at positions $u,\ldots,u+q-1$, which implies that the possible number $w$ of positions to be connected is equal to the variable node degree $q$. On the other hand, the variable nodes of degree $2$ are connected to the check nodes at the same position. We consider the SC-RA ensemble with the MET structure whose check nodes at position $v$ are connected to variable nodes of degree $q$ based on a $q\times q$ degree type matrix ${\bf S}^v$. Since $w$ is equal to $q$, ${\bf S}^v$ becomes all one matrix for all $v$. The edge connection between the variable nodes of degree $q$ and the check nodes is established using a method similar to that of the MET SC-LDPC ensemble. The remaining construction method follows the process described in Section \ref{Subsec:MET}.

The SC-RA ensemble with non-uniform degree distributions has an irregular degree distribution $\lambda_{u}(x)$ with a minimum degree of $3$ for the variable nodes in the upper side at position $u$, while the variable nodes of degree $2$ remain unchanged in the lower side. Using the design algorithm of Algorithm 3, the degree distributions of the SC-RA ensemble with non-uniform degree distributions can be obtained. Table~\ref{table:BP_SC_RA} shows that the BP threshold of the SC-RA ensemble with non-uniform degree distributions designed by Algorithm 3 is superior to that of the regular SC-RA ensemble. Note that the result obtained by Algorithm \ref{alg4} is not included because the performance improvement is insignificant compared to the results by Algorithm \ref{alg3}.

\begin{table}[t]
\centering
\caption{Comparing the BP thresholds of the regular SC-RA ensembles with the ensembles obtained by the proposed algorithm}
\label{table:BP_SC_RA}
\begin{tabular}{l|ccc}
\hline
                               & \multicolumn{3}{c}{$\epsilon^{\rm BP}$} \\
                               & $L=10$       & $L=20$      & $L=30$      \\ \hline
SC-RA, Regular & $0.5107$ & $0.4949$ & $0.4946$   \\
SC-RA, Alg. 3        & $0.5399$ & $0.5128$ & $0.5051$  \\\hline
\end{tabular}
\end{table}

\begin{figure}[t]
\centering
\subfigure[Type 1]{\includegraphics[scale=0.45]{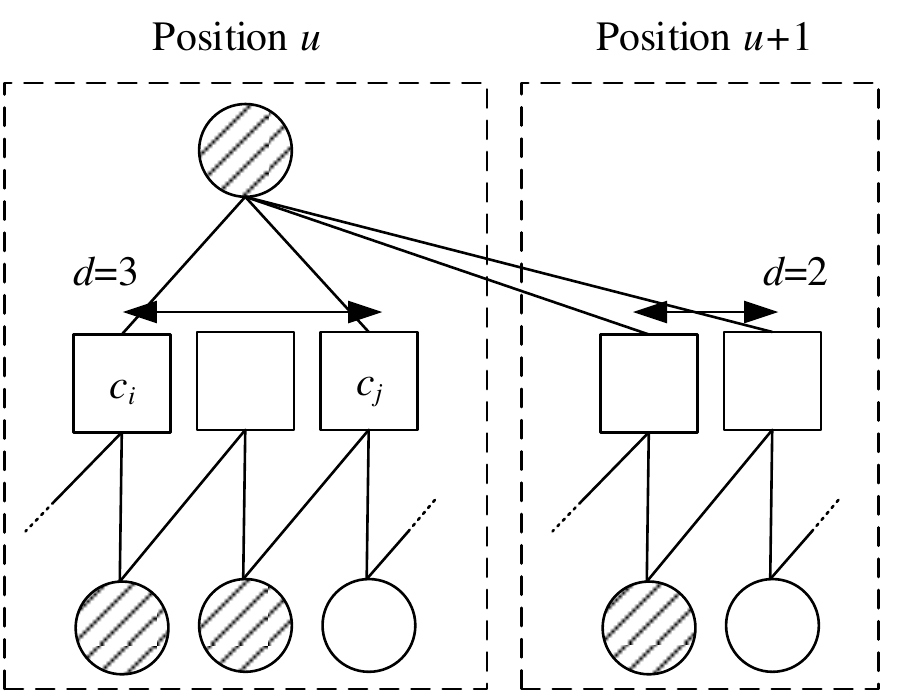}}\hspace{20pt}
\subfigure[Type 2]{\includegraphics[scale=0.45]{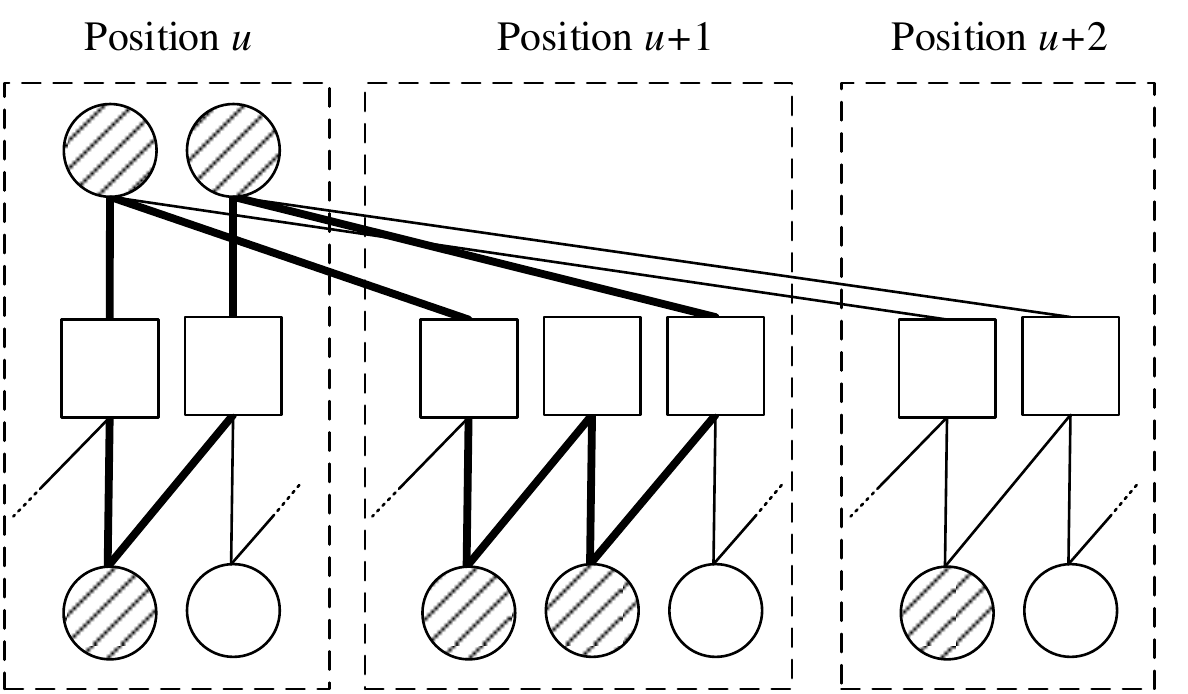}}%
\caption{Two types of stopping sets produced by low degree variable nodes.}
\label{Fig:SS}
\end{figure}

The design method of the degree distributions of the SC-RA ensemble is conceptually identical to that of the SC-LDPC ensemble. However, when we generate a code instance from the designed ensemble, small stopping sets \cite{Richardson} consisting of variable nodes of degree lower than $q$ can be made with a high probability due to the existence of variable nodes of degree 2. Therefore, it is practically important to avoid such small stopping sets for finite-length codes. We consider two types of stopping sets to be avoided, as presented in Fig.~\ref{Fig:SS}, where the stopping sets consist of the hatched variable nodes. Type $1$ stopping sets consist of one variable node of degree $i$, $3\le i<q$, with a set of variable nodes of degree $2$ and all edges of a variable node degree $i$ have at least one edge pair connected to check nodes at the same position. Similarly, type $2$ stopping sets consist of multiple variable nodes of degree $i$, $3\le i<q$, with a set of variable nodes of degree $2$ and all edges of the multiple variable nodes have at least one edge pair connected to check nodes at the same position. We disregard other stopping sets such as stopping sets consisting of variable nodes whose degrees are greater than or equal to $q$ because the probability of occurrence of such stopping sets is identical to that in the regular SC-RA ensemble.

Consider a loop consisting of $M_c$ check nodes and $M_c$ variable nodes of degree 2 at a position in a Tanner graph. If the edges between check nodes and variable nodes of degree 2 are properly connected, only one loop of length $2M_c$ is produced at each position. On the loop, define {\em distance $d$ in the loop} of two check nodes as the number of check nodes between two check nodes including themselves. For example, the distance in the loop of check nodes $c_i$ and $c_j$ in Fig.~\ref{Fig:SS}(a) is $3$.
Consider a case in which two edges of a variable node are connected to two check nodes that are separated by distance $d$ in the loop at the same position. This inevitably produces a cycle of length $2d$. Further, if all edges of a variable node are contained in such cycles, a type 1 stopping set is produced. 

However, type $1$ stopping sets are easily avoided by imposing the MET structure on variable nodes as well as check nodes.
If all edges of a variable node are connected to check nodes at different positions, type 1 stopping sets are avoided accordingly. Thus, we impose the MET structure on variable nodes such that each edge of a variable node is connected to the check node at different positions. To be specific, the first row ${\underline S}^{u,i}_1$ of the $q\times q$ degree type matrix ${\vect S}^{u,i}$ for variable nodes of degree $i$ at position $u$ is defined as
\begin{equation*}
{\underline S}^{u,i}_1=\big(\overset{i {\rm~mod~}q}{\overbrace{\lceil{i/q}\rceil,\ldots,\lceil{i/q}\rceil},}\overset{q-(i {\rm~mod~}q)}{\overbrace{\lfloor{i/q}\rfloor,\ldots,\lfloor{i/q}\rfloor}}\big)
\end{equation*}
and the remaining rows are obtained by circularly shifting the first row. For example, ${\underline S}^{u,i}_1$ for degree $i=3$ and $q=5$ is given as ${\underline S}^{u,i}_1=(1,1,1,0,0)$
which implies that all edges of variable nodes of degree $3$ are connected to check nodes located at different positions and thus type 1 stopping sets can be avoided. 

Even if all edges of a variable node are connected to check nodes at different positions, type $2$ stopping sets can be made. To avoid type $2$ stopping sets, we use the PEG algorithm \cite{Hu} with an additional constraint. Let $\mathcal{N}_{v}^{l}$ be the neighborhood of variable node $v$ within depth $l$, which denotes the set of check nodes reached by a computation tree spreading from variable node $v$ within depth $l$. In the conventional PEG algorithm, a cycle of length $2l+2$ is avoided by connecting the edges of variable node $v$ to the check nodes not in $\mathcal{N}_{v}^{l}$. However, the conventional PEG algorithm cannot prevent the production of type $2$ stopping sets if the girth of the PEG algorithm is lower than the length of the cycles in the stopping set. For example, the length of the cycle denoted by the bold line in Fig.~\ref{Fig:SS}(b) is $10$ and thus it is not avoided by the PEG algorithm with a girth such as $6$ or $8$. Thus, it is difficult to prohibit connecting the check nodes in $\mathcal{N}_{v}^{l}$  using the conventional computational tree. Instead, we consider the {\em expanded computation tree} by $d$, which is defined as follows. At depth $l$ of the conventional computation tree from variable node $v$, there exist check nodes directly connected to variable node $v$ within depth $l-1$, while the depth $l$ of the expanded tree contains not only the directly connected check nodes but also the set of check nodes within distance $d$ in the loop from the directly connected check nodes. Also, only the variable nodes of degree lower than $q$ are included without other higher degree variable nodes in the expanded computation tree because we are dealing with stopping sets consisting of such low degree variable nodes. Let $\mathcal{N}_{v}^{l,d}$ be the set of check nodes within depth $l$ in the expanded computation tree by $d$ from a variable node $v$. Then, by connecting a new edge of variable node $v$ to the check nodes not in $\mathcal{N}_{v}^{l,d}$, a cycle of a length up to $(2l+2)+(l+1)d$ is avoided in the sub-graph consisting of check nodes and low degree variable nodes. Because the expanded tree contains only low degree variable nodes, this additional constraint can be applied for large $l$ and $d$ to avoid small size of type $2$ stopping sets. For example, the cycle in Fig.~\ref{Fig:SS}(b) is avoided by prohibiting a connection between variable node $v$ and a check node in $\mathcal{N}_{v}^{1,3}$ and thus the type $2$ stopping set can be avoided. Note that because this constraint aims to avoid stopping sets consisting of low degree variable nodes, the conventional PEG algorithm should  also be applied at the same time.

\begin{figure}[t]
\centering
\subfigure[$L=10$]{\includegraphics[scale=0.30]{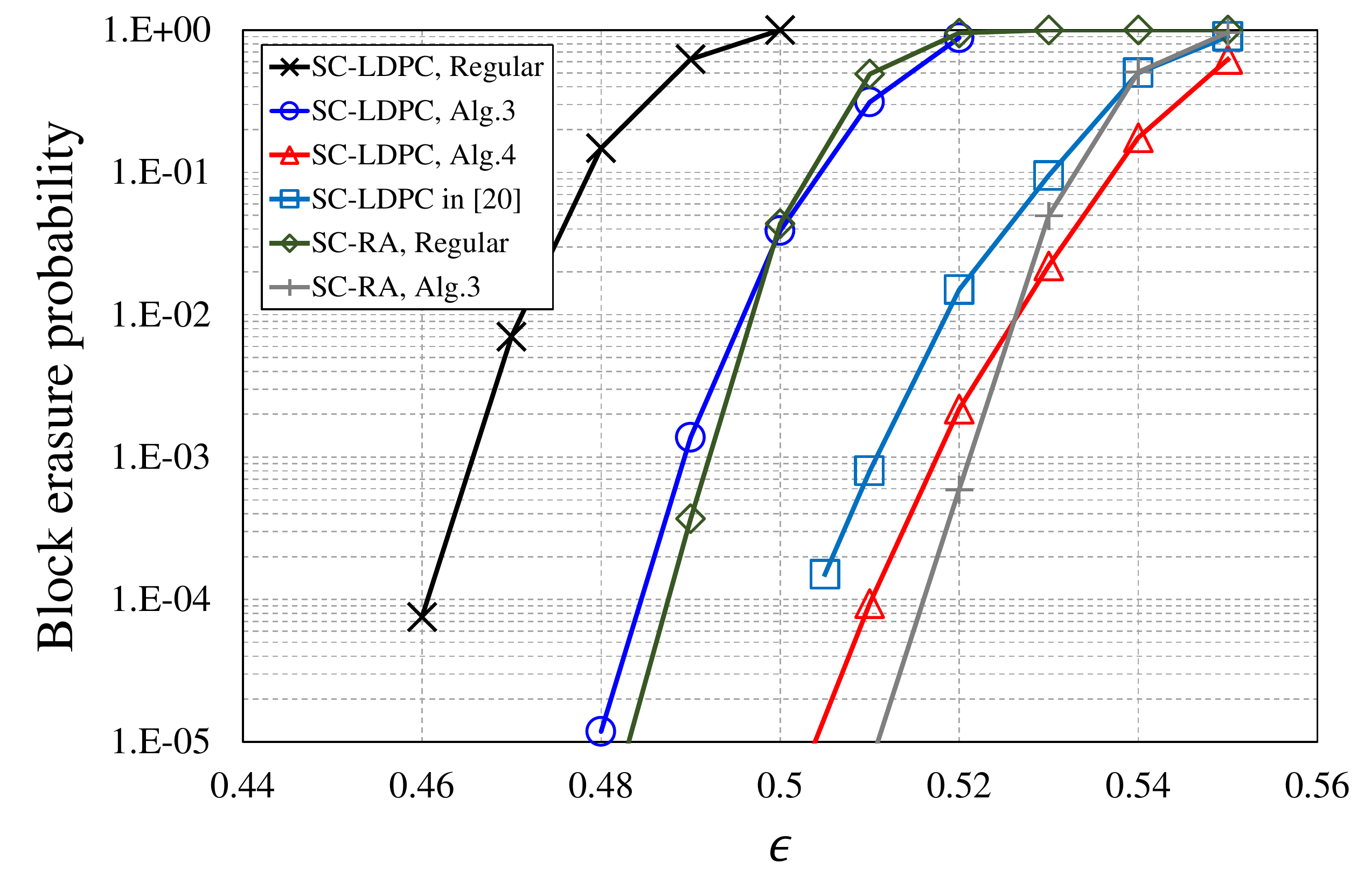}}
\subfigure[$L=20$]{\includegraphics[scale=0.30]{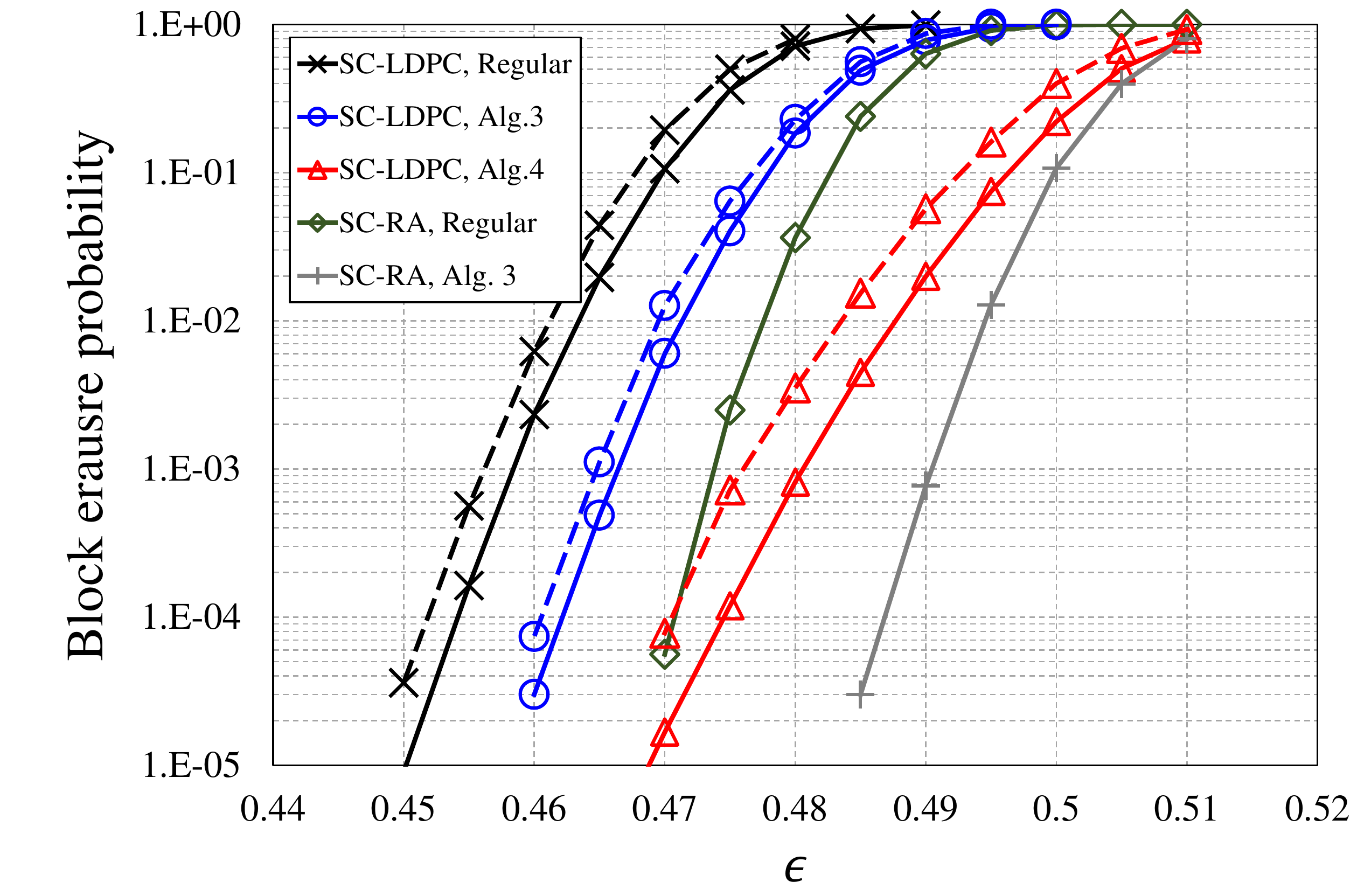}}
\subfigure[$L=30$]{\includegraphics[scale=0.30]{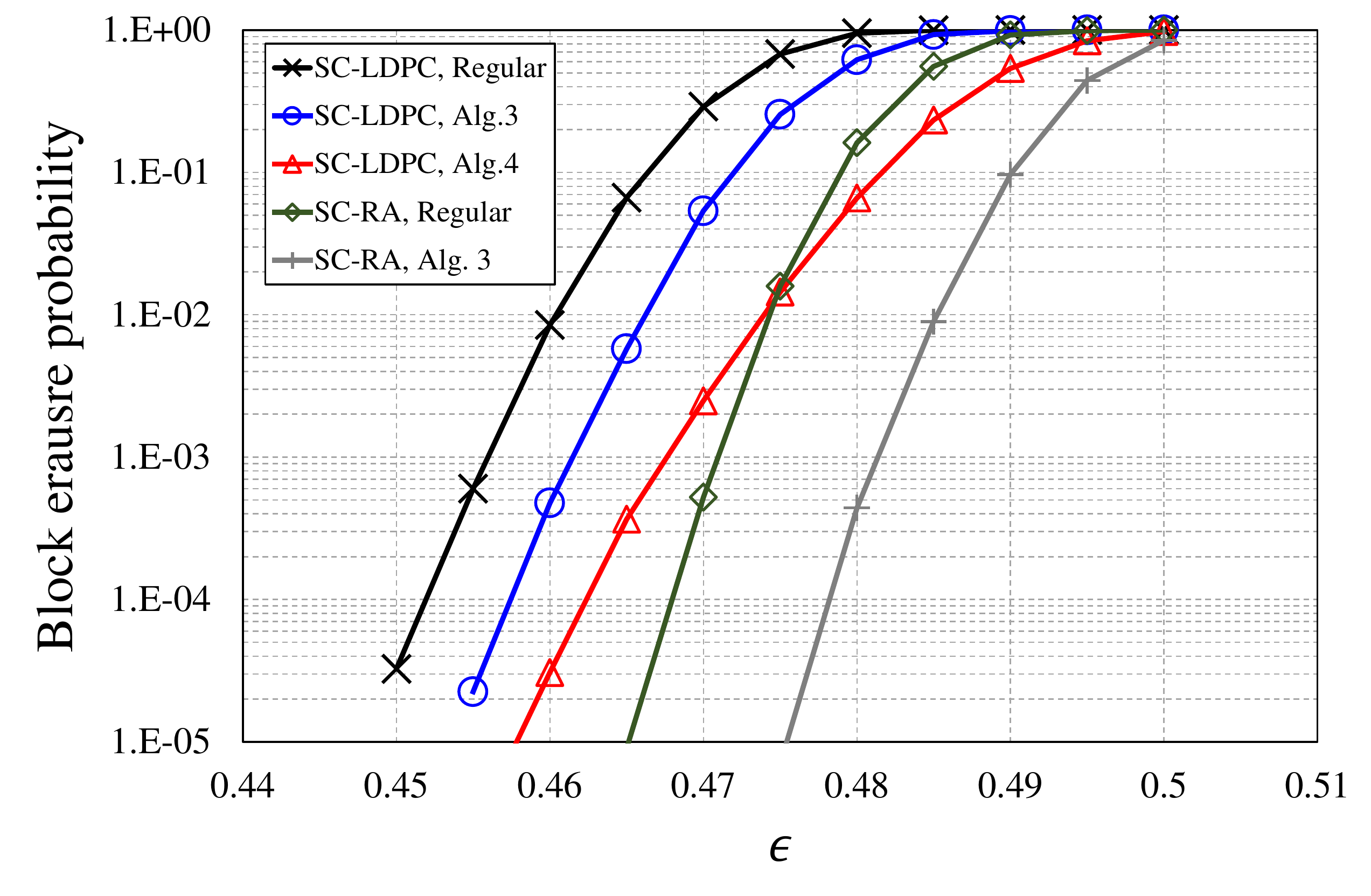}}
\caption{Block erasure probability of the SC-LDPC and SC-RA codes for $L=10$, $20$, and $30$.}
\label{Fig:FER}
\end{figure}

\subsection{Comparing Finite-Length Performances of the SC-LDPC and SC-RA Codes}

Fig.~\ref{Fig:FER} shows the block erasure probability of the SC-LDPC codes for $w=3$ and $L=10, 20, 30$. Each code instance for the SC-LDPC ensembles is obtained by the progressive edge growth (PEG) algorithm \cite{Hu} with $M=990$. The code lengths for $L=10$, $20$, and $30$ are $9,900$, $19,800$, and $29,700$, respectively. The decoder uses the BP decoding algorithm and runs the algorithm until the states of the variable nodes remain unchanged. For $L=20$ in Fig. \ref{Fig:FER} (b), the dashed lines correspond to the randomly constructed SC-LDPC codes and the solid lines correspond to the MET SC-LDPC codes. According to the simulation result, it is confirmed that the MET SC-LDPC codes have better finite-length performance than the randomly constructed SC-LDPC codes. Moreover, the improvement by Algorithms \ref{alg3} and \ref{alg4} from the regular SC-LDPC codes is observed for both the randomly constructed and the MET SC-LDPC codes. For $L=10$, we present the finite-length performances of the MET SC-LDPC codes in Fig.~\ref{Fig:FER}(a), which shows the performance improvement by the proposed algorithms. It is also shown that the SC-LDPC code obtained by Algorithm \ref{alg4} outperforms the code in \cite{Koganei} for $L=10$. The code length of the code in \cite{Koganei} is $10,000$, which is slightly larger than those of the other codes. The performance improvement by the proposed algorithms is also shown for $L=30$ in Fig.~\ref{Fig:FER}(c).

We include the simulation results for the SC-RA codes for $q=5$ in Fig.~\ref{Fig:FER}. To match their code lengths with those of the SC-LDPC codes, the SC-RA code instances are obtained from their ensembles with $M=900$. For $L=20$, the code length of all the codes in Fig.~\ref{Fig:FER}(b) is $19,800$. In terms of the code rate, the SC-RA codes have an advantage because the code rate of the SC-RA codes with $q=5$ is slightly higher than that of the SC-LDPC codes with $w=3$ for the same $L$ \cite{Stinner}, \cite{Johnson}. Each code instance of the SC-RA codes with the designed non-uniform degree distributions is obtained using the PEG algorithm with an additional constraint that avoids a connection between variable node $v$ and check nodes in $\mathcal{N}_{v}^{1,8}$. As in the SC-LDPC codes, the SC-RA codes obtained by Algorithm \ref{alg3} show better performance than the regular SC-RA codes for $L=10, 20$, and $30$. Comparing all the results in Fig.~\ref{Fig:FER}, we can confirm that the SC-RA codes obtained by Algorithm 3 are the best in terms of the finite-length performance.

\section{Conclusion}
\label{Sec: Conclusion}
In this paper, we proposed design methods for SC-LDPC codes with non-uniform degree distributions. The proposed methods are based on the local design of degree distribution at each position by solving LP problems. It was found that, if the objective function of the LP problem should be carefully selected, the designed degree distributions obtained by the proposed methods substantially improve the performance of SC-LDPC codes. We also presented the methods which improve the finite-length performance further such as imposing the MET structure and applying the design methods to the SC-RA code structure. Simulation results confirm that the proposed methods improve the finite-length performance for both SC-LDPC and SC-RA codes.

\end{document}